\documentclass[sn-mathphys]{sn-jnl}

\usepackage{graphicx,subfigure}
\graphicspath{{./figures/}} 
\usepackage{amsfonts,amsbsy,amscd,amsgen,amsthm,dsfont,amsmath,amssymb,mathrsfs,mathtools,array}
\usepackage{bm,multirow,esvect,cancel}
\theoremstyle{definition}
\usepackage{tabularx}
\usepackage{xcolor,comment,soul,url}
\usepackage{enumitem}
\usepackage[]{algorithm}
\usepackage{algpseudocode} 
\usepackage{letltxmacro}

\newcommand{\id}{\mathrm d}

\DeclareMathAlphabet\mathbfcal{OMS}{cmsy}{b}{n}

\jyear{2021}%

\theoremstyle{thmstyleone}%

\newtheorem{remark}{Remark}
\newtheorem{lemma}{Lemma}

\begin{document}

\title[Correlated uncertainties in epidemiological models]{Modeling correlated uncertainties in stochastic compartmental models}

\author[1]{\fnm{Konstantinos} \sur{Mamis}}\email{kmamis@uw.edu}
\equalcont{These authors contributed equally to this work.}

\author*[2]{\fnm{Mohammad} \sur{Farazmand}}\email{farazmand@ncsu.edu}
\equalcont{These authors contributed equally to this work.}

\affil[1]{\orgdiv{Department of Applied Mathematics}, \orgname{University of Washington}, \orgaddress{ \city{Seattle}, \postcode{98195-3925}, \state{WA}, \country{USA}}}

\affil[2]{\orgdiv{Department of Mathematics}, \orgname{North Carolina State University}, \orgaddress{\street{2311 Stinson Drive}, \city{Raleigh}, \postcode{27695-8205}, \state{NC}, \country{USA}}}

\abstract{We consider compartmental models of communicable disease with uncertain contact rates.  Stochastic fluctuations are often added to the contact rate to account for uncertainties. White noise, which is the typical choice for the fluctuations, leads to significant underestimation of the disease severity. Here,  starting from reasonable assumptions on the social behavior of individuals, we model the contacts as a Markov process which takes into account the temporal correlations present in human social activities. Consequently, we show that the mean-reverting Ornstein--Uhlenbeck (OU) process is the correct model for the stochastic contact rate. We demonstrate the implication of our model on two examples: a Susceptibles-Infected-Susceptibles (SIS) model and a Susceptibles-Exposed-Infected-Removed (SEIR) model of the COVID-19 pandemic. In particular, we observe that both compartmental models with white noise uncertainties undergo transitions that lead to the systematic underestimation of the spread of the disease. In contrast, modeling the contact rate with the OU process significantly hinders such unrealistic noise-induced transitions. For the SIS model, we derive its stationary probability density analytically, for both white and correlated noise. This allows us to give a complete description of the model's asymptotic behavior as a function of its bifurcation parameters, i.e., the basic reproduction number, noise intensity, and correlation time. For the SEIR model, where the probability density is not available in closed form, we study the transitions using Monte Carlo simulations. Our study underscores the necessity of temporal correlations in stochastic compartmental models and the need for more empirical studies that would systematically quantify such correlations.}


\maketitle

\section{Introduction}\label{sec:intro}
Compartmental models describe the spread of communicable diseases in a population \cite{Kermack1927,Capasso1978,Hethcote1994, Brauer2008,Capasso2008,Diekmann2021}.  In such models, the population of a community is partitioned into disjoint compartments, e.g., the susceptible and the infected, each one containing all individuals with the same disease status \cite{Tolles2020a}. The state variables in a compartmental model are the numbers of individuals in each compartment. The model parameters, e.g., the contact, incubation, and curing rates, determine the flow of individuals between compartments.  In the present work, we study how the model predictions are affected by the presence of uncertainties in the compartmental model parameters.

Determining the value of model parameters is a delicate task that involves estimation and averaging of data over the whole population \cite{Mummert2019,Committee2020,Ganyani2020,Zhang2020a}. As such, model parameters are subject to uncertainties, arising from the variation of social and biological factors among individuals.  Among the model parameters, average contact rate is the most volatile, due to its strong dependence on the social activity that varies from person to person,  and also changes over time \cite{Linka2020, Lloyd-Smith2005,Committee2020,Ferretti2020}. Uncertainties in the contact rate $\lambda(t)$ are often modeled as a stochastic perturbation $\xi(t)$ with intensity $\sigma$ around a constant mean $\bar{\lambda}$, so that $\lambda(t)=\bar\lambda+\sigma\xi(t)$. A common choice for $\xi(t)$ is Gaussian white noise, see e.g. \cite{Gray2011}. The remaining model parameters, such as the average incubation or curing rate, depend mainly on the biology of the virus, and, while every individual responds to the infection differently, they vary less compared to the contact rate and can be considered constant.

In a recent study \cite{Mamis2022}, we studied the role of temporal correlations, which are present in social activities of individuals, on the contact rate $\lambda(t)$. Using standard results from the theory of stochastic processes, and assuming that the perturbation $\xi(t)$ has an exponentially decreasing autocorrelation function, we showed that the only admissible model for the stochastic contact rate is the Ornstein--Uhlenbeck (OU) process. However, the assumption of exponentially decreasing autocorrelation had remained unjustified in Ref.~\cite{Mamis2022}.
A main contribution of the present paper is to derive the OU process without making such an onerous assumption. In fact, as we show in Sec.~\ref{sec:OU_justification}, the OU process emerges naturally by making quite simple and realistic assumptions on the contacts of each individual in the population.

Then, we focus on determining the final size of disease in a population, as predicted by compartmental models with white or OU noise fluctuations in contact rate. We study two compartmental models. First, we consider the stochastic Susceptibles-Infected-Susceptibles (SIS) model, which is adequate for modeling sexually transmitted or bacterial diseases such as gonorrhea or syphilis. For the SIS model, we determine the stationary probability density (PDF) of the infected population fraction in closed form. This allows us to completely classify the asymptotic state of the model as a function of its bifurcation parameters, i.e., basic reproduction number, noise intensity, and the correlation time of the noise. As the second model, we consider the stochastic Susceptibles-Exposed-Infected-Removed (SEIR) model for COVID-19 pandemic in the US during the Omicron variant. Using Monte Carlo simulations, we study the bifurcations of the asymptotic probability density as in the SIS model.

Our main qualitative result is that, for increasing levels of white noise in the contact rate, both compartmental models undergo a noise-induced transition, whereby the stationary PDF of the infected exhibits an additional peak near zero, and far away from the deterministic equilibrium. This unrealistic behavior leads to significant underestimation of the severity of the disease.
In contrast, under OU noise this transition is suppressed, with most of the probability mass of the stationary PDF being concentrated around the deterministic equilibrium. 

\subsection{Related work}\label{sec:related_works}
Stochasticity has been incorporated into many epidemiological models \cite{Andersson2000,Allen2010}. One principled approach for deriving stochastic compartmental models begins by modeling of the number of individuals in each compartment as continuous-time Markov chain birth–death processes or as branching processes, see e.g.~\cite{Allen2017, Black2010,Britton2010}. Then, by assuming their state variables to be continuous, the Markov chain models result in a system of stochastic differential equations (SDEs) whose parameters contain white noise uncertainties. Another approach is to directly add noise to the model parameters; see e.g. \cite{Faranda2020a, Gray2011,Maki2013,Linka2020,Neri2021}. This approach is more straightforward, but the choice of the type of noise is largely arbitrary: The most common choice in literature is Gaussian white noise \cite{Gray2011,Ji2014,Cai2015,Meng2016,Cai2020}. However, OU noise has also been proposed for parameter perturbation in biological systems, see e.g., \cite{Liu2022,Wang2018,Bartoszek2017,Allen2016,Rohlfs2014,Gray2011,Aalen2004}, because OU noise combines the modeling of stochastic fluctuations with the stabilization around an equilibrium point, due to its mean-reverting property.

Recently, the COVID-19 pandemic has renewed interest in stochastic modeling of disease spread; see e.g.~\cite{Bertozzi2020} for a survey of existing forecast models for COVID-19, and~\cite{Faranda2020a} where a lognormally distributed process has been considered for the stochastic fluctuations in contact rate of a COVID-19 compartmental model, to account for the presence of superspreaders in the population. However, most of the studies that use stochastic compartmental models to make predictions for the COVID-19 pandemic rely primarily on simulations, see e.g., \cite{Neri2021,Faranda2020,Faranda2020a}.  

Most of the analysis performed on stochastic compartmental models has been focused on the derivation of conditions for the eradication or persistence of the disease in the population; see e.g., \cite{Cai2020,Meng2016,Cai2015,Ji2014,Zhao2014,Lahrouz2013,Gray2011,Tornatore2005,Li2004} for compartmental models with white noise uncertainties. Recently, this line of work has been extended to models with OU uncertainties \cite{Liu2022,Wang2018}, and L\'evy noises \cite{Chen2017a,Zhang2013} to account for abrupt changes (jumps) in disease transmission. 

The focus of the present work is different; apart from conditions for the disease to become endemic, we are also interested in the predictions of stochastic compartmental models on the final disease size.  Analytic work in this direction is scarce; see e.g. Ref. \cite{Mendez2012} which uses the Fokker--Planck equation to study a scalar compartmental model under white noise fluctuations in contact rate.

\subsection{Outline}\label{sec:outline}
This paper is organized as follows. In Sec.~\ref{sec:OU_justification}, we present our model of uncertainties in the contact rate. In Sec.~\ref{sec:det_SIS}, we study the SIS model, for both cases of white and OU noise fluctuations in contact rate. We analytically determine the noise-induced transitions the stochastic SIS model undergoes, and we quantify the effect of noise correlations in contact rate. In Sec.~\ref{sec:seir}, we study the noise-induced transitions of a stochastic SEIR model for the Omicron wave of the COVID-19 pandemic in the US, by using direct Monte Carlo simulations. In Sec.~\ref{sec:conclusions}, we make our concluding remarks and outline possible directions for future work.

\section{Modeling uncertainties in contact rate}\label{sec:OU_justification}
The average contact rate $\lambda$, defined as the average number of adequate contacts per individual per unit time \cite[Sec. 2.7.1]{Vynnycky2010}, \cite{Tolles2020}, is the main source of uncertainty in compartmental models. In order to determine its properties as a random process, we begin with the cumulative number of contacts $C_n(t)$ of the $n$-th individual up to time $t$. We denote the incremental number of contacts by $\Delta C_n(t)$ which measures the number of constants that the $n$-th individual makes during the time interval $[t,t+\Delta t]$, with $\Delta t$ being a reference time interval, e.g., a day or a week. The contact rate $\lambda_n$ of the $n$-th individual is then given by $\lambda_n(t) = \Delta C_n(t)/\Delta t$.

Next, we make the following assumptions on the social behavior of each individual.
\begin{enumerate}
\item The average number of contacts that individuals make in a time interval is proportional to the length of the time interval, so that
$\mathsf E[\Delta C_n(t)]=\mu_n \Delta t,$ for some constant $\mu_n>0$. 
\item The number of contacts $\Delta C_n(t)$ is subject to time-varying random fluctuations, the intensity of which is also proportional to reference unit time $\Delta t$. For instance, this assumption implies that the contacts of an individual per week are prone to more uncertainty than the contacts of the same individual per day.
\item After a period of relatively high or low contacts compared to the average number $\mu_n\Delta t$, the contacts of the individual will tend towards the mean $\mu_n\Delta t$. In other words, high or low numbers of contacts are not sustained for prolonged periods of time. 
\end{enumerate}
\begin{figure}
\centering
\includegraphics[width=.95\textwidth]{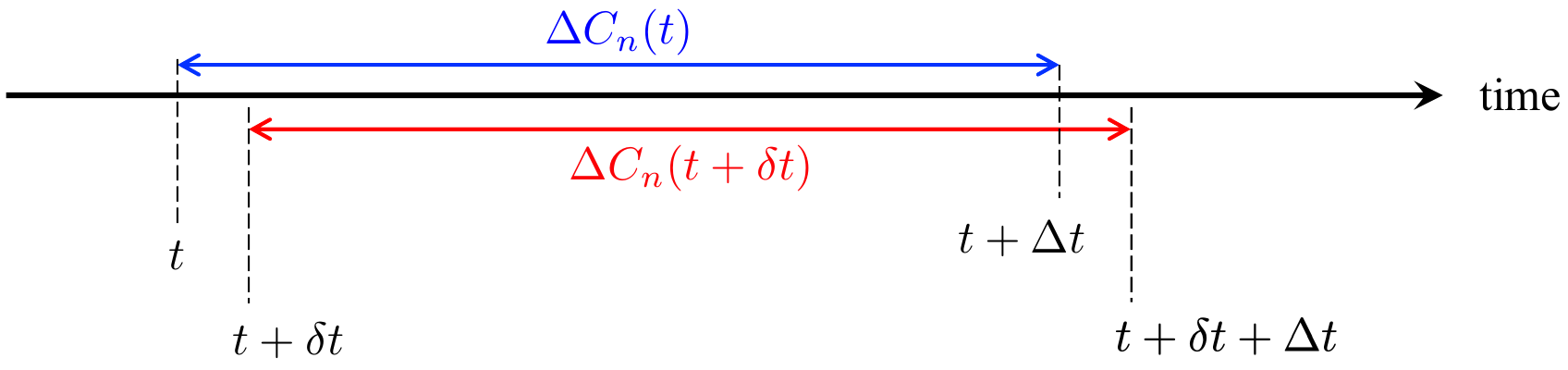}
\caption{Illustration of the quantities in the conditional probability~\eqref{eq:transition}.}
\label{fig:schem}
\end{figure}
Under the above assumptions, we formulate the conditional probability of $\Delta C_n(t+\delta t)-\Delta C_n(t)$, given the number of contacts $\Delta C_n(t)$, where $\delta t\ll \Delta t$ is a small time increment. Note that $\Delta C_n(t+\delta t)$ is the number of contacts that the individual makes over the time interval $[t+\delta t,t+\delta t+\Delta t]$. Therefore, $\Delta C_n(t+\delta t)-\Delta C_n(t)$ measures the variations in the number of contacts as the reference time interval $[t,t+\Delta t]$ is shifted ever so slightly (see Fig.~\ref{fig:schem} for an illustration).

Based on the above assumptions, for given positive integers $i$ and $j$, we define the conditional probability,
\begin{align}\label{eq:transition}
\mathsf{P}\big[\Delta C_n(t+\delta t)&-\Delta C_n(t)=j\vert\Delta C_n(t)=i\big]=\nonumber\\
&
\begin{cases}
\frac{1}{2}[(\kappa_n \Delta t)^2-\theta_n(\mu_n \Delta t-i)]\delta t & j=-1\\
\frac{1}{2}[(\kappa_n \Delta t)^2+\theta_n(\mu_n \Delta t-i)]\delta t & j=+1\\
1-(\kappa_n \Delta t)^2\delta t & j=0\\
0 & \text{otherwise},
\end{cases}
\end{align} 
where $\kappa_n$, $\theta_n$ and $\mu_n$ are positive constants. We will see shortly that $\kappa_n$ controls the noise intensity and $\theta_n$ determines the time correlation 
of the resulting stochastic process. For simplicity, we assume that these constants are identical across the population. Therefore, we omit the subscript $n$ and simply denote them by $\kappa$, $\theta$ and $\mu$.

The conditional probability~\eqref{eq:transition} dictates the following. If the current number of contacts $\Delta C_n(t) = i$ is greater than the mean $\mu\Delta t$, then it is more probable for the number of contact to decrease by one after a short time interval $\delta t$ has passed (case $j=-1$). Conversely, if the number of contacts $\Delta C_n(t) = i$ is less than the mean, it is more likely for the number of contacts to increase by one in the near future (case $j=+1$). Furthermore, it assumes that the probability that the number of contacts jump by more than one within a short time $\delta t$ is negligible. Finally, the probability for case $j=0$ (no change within time $\delta t$) is defined to ensure that the total probability adds up to one. 

We note that the constant $\theta$ plays a crucial role here. If $\theta =0$, the number of contacts increase or decrease with the same probability and regardless of their past history (Brownian motion). In contrast, $\theta >0$ introduces time correlations into the process so that the number of contacts have a tendency to revert back to their mean value.

From Eq.~\eqref{eq:transition}, we calculate the conditional mean value and variance,
\begin{subequations}\label{eq:EVar}
\begin{equation}
\mathsf{E}[\Delta C_n(t+\delta t)-\Delta C_n(t)\vert\Delta C_n(t)=i]=\theta(\mu \Delta t-i)\delta t, 
\end{equation}
\begin{equation}
\mathsf{Var}[\Delta C_n(t+\delta t)-\Delta C_n(t)\vert\Delta C_n(t)=i]=(\kappa \Delta t)^2\delta t.
\end{equation}
\end{subequations}
Using Eq.~\eqref{eq:EVar}, and the definition $\lambda_n(t) = \Delta C_n(t)/\Delta t$, we calculate the conditional mean and variance,
\begin{subequations}\label{eq:EVar_2}
\begin{align}
\mathsf{E}[\lambda_n(t+\delta t)&-\lambda_n(t)\vert\lambda_n(t)=\alpha]=  \nonumber\\
&\frac{1}{\Delta t}\mathsf{E}[\Delta C_n(t+\delta t)-\Delta C_n(t)\vert\Delta C_n(t)=\alpha\Delta t]=\theta(\mu-\alpha)\delta t, 
\end{align}
\begin{align}
\mathsf{Var}[\lambda_n(t+\delta t)&-\lambda_n(t)\vert\lambda_n(t)=\alpha]=\nonumber\\
&\frac{1}{(\Delta t)^2}\mathsf{Var}[\Delta C_n(t+\delta t)-\Delta C_n(t)\vert\Delta C_n(t)=\alpha\Delta t]=\kappa^2\delta t,
\end{align}
\end{subequations}
where $\alpha=i/\Delta t$.
Assuming no dependence between the incremental contacts of different individuals, $\lambda_n(t)$ are independent random variables. This means that $\lambda_n(t+\delta t)-\lambda_n(t)$ are also independent random variables, with the same mean value and variance given by Eq.~\eqref{eq:EVar_2}. Hence, the central limit theorem implies that the average over the whole population of $N$ individuals,
\begin{equation}
\lambda(t+\delta t)-\lambda(t)=\frac{1}{N}\sum_{n=1}^N\left(\lambda_n(t+\delta t)-\lambda_n(t)\right),
\end{equation}
follows a normal distribution with mean $\theta(\mu-\alpha)\delta t$ and variance $\kappa^2\delta t/N$. As a result, we have
\begin{equation}\label{eq:clt0}
\lambda(t+\delta t)-\lambda(t)=\theta(\mu-\alpha)\delta t+D\sqrt{\delta t}\mathcal N(t),
\end{equation}
where $D=\kappa/\sqrt{N}$, and $\mathcal N(t)$ is the standard normal distribution, with $\mathcal{N}(t)$ and $\mathcal{N}(s)$ being independent for $t\neq s$. 
Recall that the expressions in Eq.~\eqref{eq:EVar_2} are conditioned on $\lambda_n(t)=\alpha$ for all $n=1,\ldots,N$, which implies $\lambda(t)=(1/N)\sum_{n=1}^N\lambda_n(t)=\alpha$. Therefore Eq.~\eqref{eq:clt0} is equivalent to
\begin{equation}\label{eq:clt}
\lambda(t+\delta t)-\lambda(t)=\theta(\mu-\lambda(t))\delta t+D\sqrt{\delta t}\mathcal N(t).
\end{equation}
Dividing by $\delta t$ and taking the limit $\delta t\to 0$, we obtain the Langevin equation,
\begin{equation}
	\frac{\id\lambda(t)}{\id t} = \theta (\mu - \lambda(t)) + D\xi^{\mathrm{WN}}(t),
	\label{eq:OU_SDE}
\end{equation}
where $\xi^{\mathrm{WN}}(t)$ is the standard white noise.
Equation~\eqref{eq:OU_SDE} is the SDE for an Ornstein--Uhlenbeck process. Therefore, the average contact rate $\lambda(t)$ is an OU process. 

The stationary solution of Eq.~\eqref{eq:OU_SDE} is a Gaussian process with the following mean and autocovariance~\cite{Hanggi1995},
\begin{equation}\label{eq:m_cov_1}
\mathsf{E}[\lambda(t)]=\mu, \ \ \mathsf{Cov}[\lambda(t)\lambda(s)]=\frac{D^2}{2\theta}\exp\left(-\theta\vert t-s\vert \right).
\end{equation}
Introducing the new parameters $\tau=1/\theta$, $\sigma^2=D^2\tau^2$, mean value and autocovariance of Eq.~\eqref{eq:m_cov_1} are recast into
\begin{equation}\label{eq:m_cov_2}
\mathsf{E}[\lambda(t)]=\mu, \ \ \mathsf{Cov}[\lambda(t)\lambda(s)]=\frac{\sigma^2}{2\tau}\exp\left(-\frac{\vert t-s\vert}{\tau}\right).
\end{equation}
Now it is easy to see that $\tau=1/\theta$ is the correlation time of the average contact rate $\lambda(t)$. It can be shown that, as $\tau\rightarrow 0$, the autocovariance~\eqref{eq:m_cov_2} tends to the delta function, corresponding to white noise with intensity $\sigma$~\cite[Sec. 6.6]{Toral2014}.

We note that the autocovariance in Eq.~\eqref{eq:m_cov_2} was assumed in the earlier derivation of Mamis and Farazmand~\cite{Mamis2022}. Here, we have shown that this property can be deduced naturally from the conditional probability~\eqref{eq:transition}.

In the following sections, to simplify the notation, we write $\lambda(t)=\bar{\lambda}+\sigma\xi^{\mathrm{OU}}(t)$, where $\bar{\lambda}=\mu$ is the mean value, $\sigma$ is the noise intensity, and $\xi^{\mathrm{OU}}(t)$ is the standard OU process. The standard OU process $\xi^{\mathrm{OU}}(t)$ has zero mean and its autocovariance is given by 
\begin{equation}\label{eq:OU_cor}
\mathsf{E}\left[\xi^{\text{OU}}(t)\xi^{\text{OU}}(s)\right]=\frac{1}{2\tau}\exp\left(-\frac{\vert t-s\vert}{\tau}\right).
\end{equation} 
With this expression, $\lambda(t) =\bar{\lambda}+\sigma\xi^{\mathrm{OU}}(t)$ satisfies the Langevin Eq.~\eqref{eq:OU_SDE} and its mean and covariance are given by Eq.~\eqref{eq:m_cov_2}.

\section{SIS model}\label{sec:det_SIS}
The Susceptibles-Infected-Susceptibles (SIS) model is described by the equations
\begin{subequations}\label{eq:SIS_2}
	\begin{equation}\label{eq:S}
		\frac{\id S(t)}{\id t}=-\frac{\lambda}{N}S(t)I(t)+\gamma I(t),
	\end{equation}
	\begin{equation}\label{eq:I}
		\frac{\id I(t)}{\id t}=\frac{\lambda}{N}S(t)I(t)-\gamma I(t),
	\end{equation}
\end{subequations}
where $S(t)$, $I(t)$ are the numbers of susceptible and infected individuals, respectively, and $N$ is the total population. SIS model parameters are the average contact rate $\lambda$ and the average curing rate $\gamma$, which is the inverse of the average time an individual needs to recover.  SIS Eq.~\eqref{eq:SIS_2} is suitable for modeling diseases that are curable, and whose infection does not confer protective immunity; thus the infected become susceptibles again after their recovery.  This is the case for most  bacterial and sexually transmitted diseases~\cite{Liljeros2003, Hethcote1984}.  

Note that $(\lambda/N)S(t)I(t)$ is the simplest form for the disease transmission term, and it is based on the assumption of homogeneous mixing of population \cite{Tolles2020a}. Under this assumption, out of the total number of contacts that each susceptible individual makes on average per unit time,  $\lambda I/N$ contacts are with the infected, resulting in disease transmission. Transmission term without the division with $N$ is sometimes used \cite{Gray2011,Mendez2012}; however, this choice is not supported by empirical evidence \cite{Rhodes2008}.

Under the usual assumption of constant population $S(t)+I(t)=N$, SIS model \eqref{eq:SIS_2} can be reduced to one scalar ordinary differential equation (ODE) \cite{Gray2011}. Defining the infected fraction of the population, $X(t)=I(t)/N\in[0,1]$, as the state variable, the scalar ODE is written as
\begin{equation}\label{eq:SIS}
\frac{\id X(t)}{\id t}=\lambda X(t)(1-X(t))-\gamma X(t).
\end{equation}
The equilibrium points of ODE \eqref{eq:SIS} are $x_0=0$, and $x_1=(\lambda-\gamma)/\lambda$.  The  stability of equilibrium points depends on the basic reproduction number $R_0=\lambda/\gamma$:
\begin{itemize}
\item For $R_0<1$, equilibrium point $x_0=0$ is stable. In this case, the disease is eventually eradicated from the population.
\item For $R_0>1$, equilibrium point $x_0=0$ is unstable and $x_1=(\lambda-\gamma)/\lambda$ is stable. In this case, the disease persists in the population and becomes endemic.
\end{itemize}
In the endemic case, $R_0>1$, we derive a characteristic time scale for ODE \eqref{eq:SIS}. For this, we linearize ODE \eqref{eq:SIS} around the stable equilibrium $x_1$,  and calculate its Lyapunov exponent $\lambda-\gamma$ (see also \cite[Sec. V.A]{Mamis2021}). The characteristic time scale is determined as the inverse of the Lyapunov exponent, $\eta=(\lambda-\gamma)^{-1}$.

Under the stochastic perturbation of the contact rate $\lambda(t)=\bar\lambda+\sigma\xi(t)$, the SIS model reads
\begin{equation}\label{eq:SIS_stoch}
\frac{\id X(t)}{\id t}=\bar{\lambda} X(t)(1-X(t))-\gamma X(t)+\sigma X(t)(1-X(t))\xi(t).
\end{equation}
Eq.~\eqref{eq:SIS_stoch} is a stochastic differential equation under multiplicative noise excitation, since noise excitation $\xi(t)$ is multiplied by a state-dependent function. 

In the remainder of this section, we determine the asymptotic behavior of SIS model \eqref{eq:SIS_stoch} for two cases: 1. when $\xi(t)$ is the standard Gaussian white noise, and 2. when $\xi(t)$ is the standard OU process. In particular, we show that the OU process, as derived in Sec.~\ref{sec:OU_justification}, is more suitable for modeling uncertainties in the contact rate.
In contrast to the deterministic SIS model \eqref{eq:SIS}, stochastic SIS model \eqref{eq:SIS_stoch} exhibits a richer asymptotic behavior that includes regions of bistablity,  and regions with $R_0>1$ where $x_0=0$ is stable.  

For the SIS model under white noise, its stationary PDF is easily determined as the stationary solution of the classical Fokker--Planck equation, see, e.g., \cite[Ch. 5]{Gardiner2004}.  For the case of OU fluctuations in contact rate, determining the stationary PDF is not as straightforward; the derivation and solution of Fokker--Planck-like equations, corresponding to stochastic differential equations (SDEs) excited by correlated noise, has been the topic of research for many decades \cite{Sancho1982, Hanggi1985, Fox1986, Peacock-Lopez1988,Hanggi1995,Ridolfi2011,Venturi2012, Bianucci2020a}.  Recently \cite{Mamis2021,Mamis2019a},  we have proposed a nonlinear Fokker--Planck equation whose validity is not limited to small correlation times of the stochastic excitation (see Appendix \ref{A:FP}). As we show in Sec. \ref{sec:SIS_OU}, the stationary solution to this nonlinear Fokker--Planck equation is given in explicit closed form for the case of the stochastic SIS model.  Thus, stochastic SIS model under OU perturbation is a rare instance of a nonlinear SDE under correlated noise whose stationary solution can be analytically determined. 

By having the stationary PDFs in explicit form for both white and OU models, we are able to systematically investigate the noise-induced transitions that the stochastic SIS model undergoes, for increasing levels of noise.

\subsection{SIS model under white noise}
In this section, we consider $\xi(t)$ to be the standard white noise $\xi^{\text{WN}}(t)$ with zero mean value and autocorrelation
\begin{equation}\label{eq:WN_cor}
\mathsf{E}\left[\xi^{\text{WN}}(t)\xi^{\text{WN}}(s)\right]=\delta(t-s), 
\end{equation}
where $\mathsf{E}[\cdot]$ denotes the expected value and $\delta(t-s)$ is Dirac's delta function.
For the stochastic SIS model \eqref{eq:SIS_stoch} under white noise, we calculate the stationary PDF of $X(t)$ as the stationary solution to the corresponding Fokker--Planck equation (see Appendix \ref{A:FP}),
\begin{equation}\label{eq:p0_white}
p_0(x)=Cx^{\frac{2\left(1-R_0^{-1}\right)}{\left(\sigma^2/\bar{\lambda}\right)}-2+\varpi}(1-x)^{-\frac{2\left(1-R_0^{-1}\right)}{\left(\sigma^2/\bar{\lambda}\right)}-2+\varpi}\exp\left(-\frac{2R_0^{-1}}{\left(\sigma^2/\bar{\lambda}\right)}\frac{1}{1-x}\right),
\end{equation}
where $C$ is a normalization factor, so that $\int_{\mathbb{R}}p_0(x)\id x=1$.  Parameter $\varpi$ models the difference, on the level of stationary PDF, between the It\=o ($\varpi=0$) and Stratonovich ($\varpi=1$) solution of SDE \eqref{eq:SIS_stoch} under white noise. This difference stems from the different definition of integrals with respect to Wiener process in the two approaches \cite{Arnold1974}.  
\begin{figure}
	\centering
	\includegraphics[width=\textwidth]{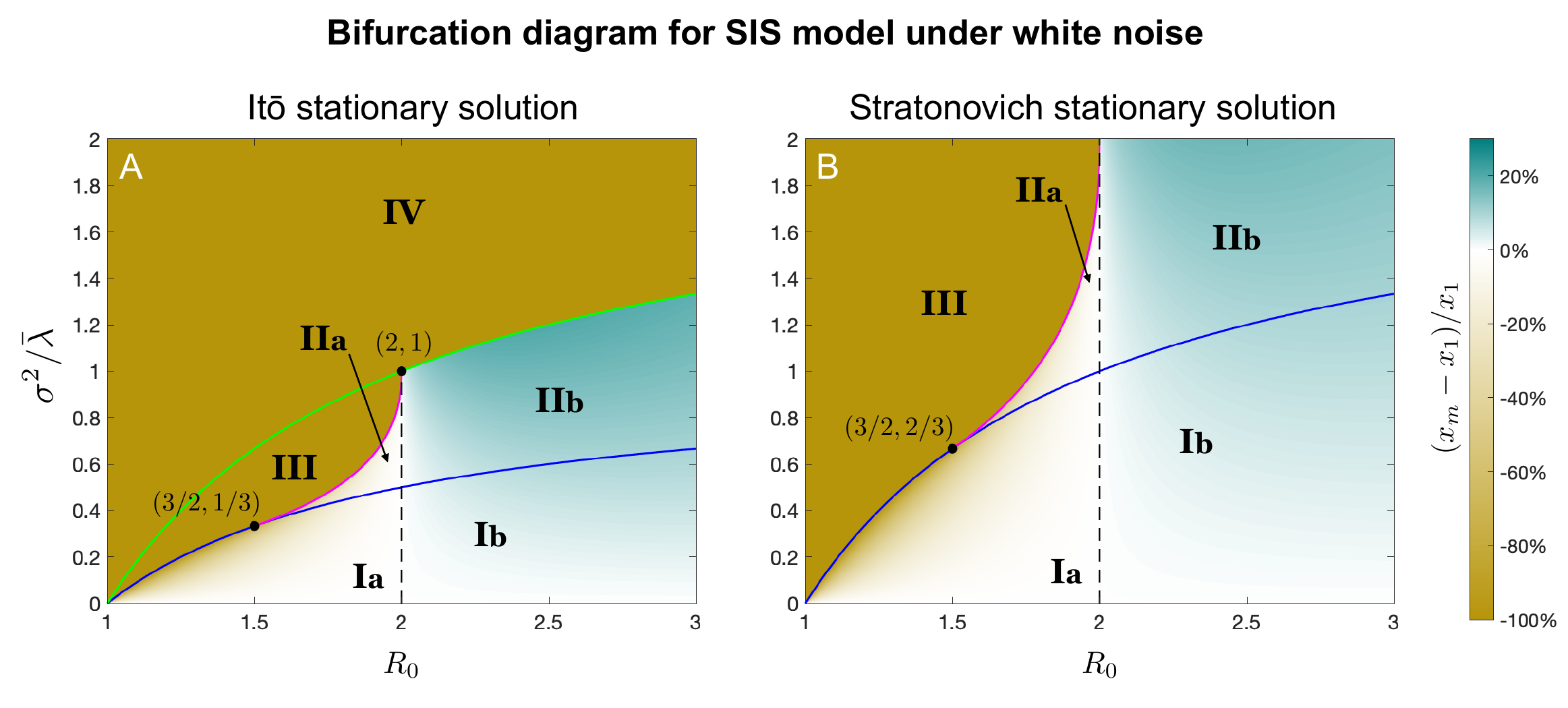}
	\caption{Bifurcation diagrams for the shape of stationary PDF of infected population fraction of stochastic SIS model \eqref{eq:SIS_stoch} under white noise,  for It\=o \textit{(panel A)} and Stratonovich \textit{(panel B)} solutions.  We define the depicted curves in Appendix \ref{A:extrema}.}\label{fig:bifurcation_white}
\end{figure}

Stationary PDF \eqref{eq:p0_white} depends on two dimensionless parameters: the basic reproduction number of the underlying deterministic model, $R_0=\bar{\lambda}/\gamma$, and the relative variance of the noise, $\sigma^2/\bar{\lambda}$, measuring the noise intensity.  Using these dimensionless parameters, we study the bifurcation diagram of the stationary PDF as shown in Fig.~\ref{fig:bifurcation_white} (see Appendix \ref{A:extrema} for calculations). As we derive in Appendix \ref{A:extrema}, both It\=o and Stratonovich solutions result in a disease eradication for $R_0<1$, as is the case for the deterministic SIS model. Therefore, we only consider the range $R_0>1$ in Fig.~\ref{fig:bifurcation_white}. 

The different regions in Fig.~\ref{fig:bifurcation_white}, marked by roman numerals, correspond to different shapes of the stationary PDF of the infected population fraction $X$: 
\begin{enumerate}[label=(\Roman*)]
\item Unimodal with mode at a non-zero $x_m$: The most probable outcome is the disease to become endemic in the population.
\item Bimodal with one mode at zero and one at a non-zero $x_m$: In this case, the most probable outcomes is either the disease being eradicated, or to attain the level $x_m$ in the population. 
\item Unimodal with mode at zero: The disease is most likely eradicated from the population.
\item Delta function at zero,  present only for It\=o solution: disease eradication is certain.  This is the only case of absolute eradication of the disease for $R_0>1$.  In the It\=o solution, the disease persists in the population for $(\sigma^2/\bar{\lambda})<2\left(1-1/R_0\right)$ (region below the green curve in Fig.~\ref{fig:bifurcation_white}A),  written equivalently as
\begin{equation}\label{eq:gray}
R_0-\frac{\sigma^2}{2\gamma}>1.
\end{equation}
Eq.~\eqref{eq:gray} is the disease persistence condition derived in \cite{Gray2011}, as expressed for the stochastic SIS model \eqref{eq:SIS_stoch}.  Thus, for the It\=o solution, increase in noise intensity results eventually in the eradication of the disease from the population, regardless of the value of $R_0$. On the other hand, region IV is absent from Fig.~\ref{fig:bifurcation_white}B, meaning that, under Stratonovich interpretation, the disease is never surely eradicated from the population for $R_0>1$, see also~\cite{Mendez2012}.
\end{enumerate}

\begin{figure}
	\centering
	\includegraphics[width=\textwidth]{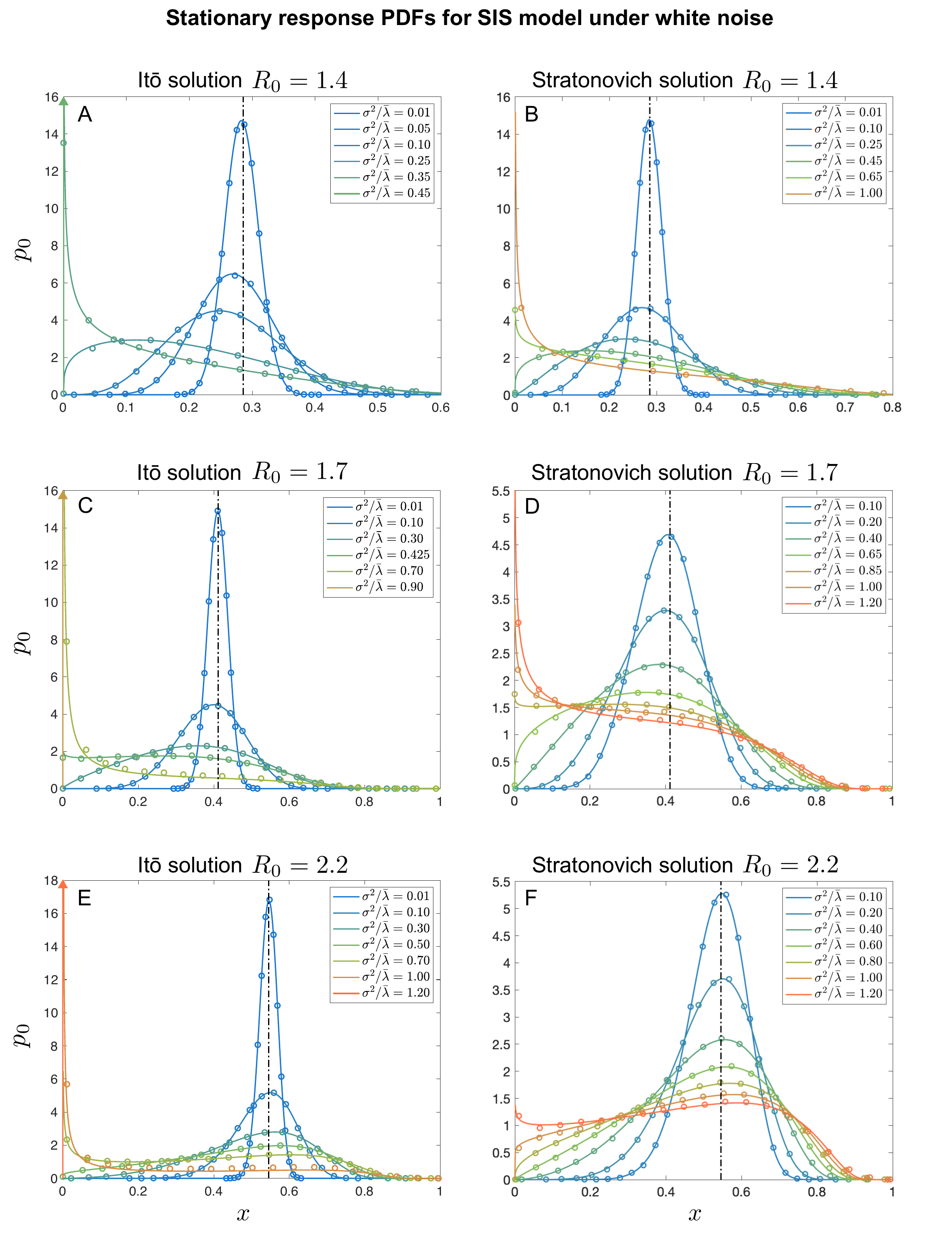}
	\caption{Stationary PDFs of infected population fraction of stochastic SIS model \eqref{eq:SIS_stoch} under white noise, for It\=o \textit{(left column)} and Stratonovich \textit{(right column)} solution. In every figure, the deterministic equilibrium is denoted by a dash-dotted line,  the open circles correspond to PDFs obtained from 50,000 samples of direct Monte Carlo simulations, while solid lines mark the PDFs obtained from the analytic solution \eqref{eq:p0_white}.  In cases where PDF is a delta function at zero, we depict it as a vertical arrow.}\label{fig:pdfs_white}
\end{figure}

Apart from the PDF shape, another important measure of disease severity is the value of $x_m$, at which the non-zero PDF mode is exhibited.  As we observe in Fig. \ref{fig:pdfs_white}, for low levels of noise, the stationary PDFs of the infected population fraction are narrow and unimodal, exhibiting their mode at the stable equilibrium $x_1$ of the underlying deterministic SIS model.  As the noise level increases, the PDF mode $x_m$ moves away from the deterministic equilibrium. This phenomenon is called the \textit{peak drift} \cite{Hanggi1995,Mamis2019a,Mamis2021}, and is commonplace in SDEs with multiplicative noise excitation such as SDE~\eqref{eq:SIS_stoch}.  

The color in Fig.~\ref{fig:bifurcation_white} encodes the peak drift phenomenon quantifying the difference between the coordinates $x_m$ (non-zero PDF mode) and $x_1$ (deterministic equilibrium point), as a percentage of $x_1$.  Figure~\ref{fig:bifurcation_white} revels two opposite trends in peak drift. In regions Ia and IIa to the left of the vertical dashed line where $R_0<2$, higher noise intensity $\sigma^2/\bar{\lambda}$ results in the non-zero PDF mode $x_m$ to drift towards zero.
In contrast, in regions Ib and IIb where $R_0>2$, higher noise intensity $\sigma^2/\bar{\lambda}$ results in the non-zero PDF mode $x_m$ to drift towards one.

The above discussion shows that, by increasing the relative noise intensity $\sigma^2/\bar{\lambda}$, the stochastic SIS model undergoes a \textit{noise-induced transition} \cite{Ridolfi2011}, i.e., a bifurcation in the shape of its stationary PDF. The type of noise-induced transition is determined by the value of the deterministic dimensionless parameter $R_0$:
\begin{itemize}
\item \textit{Type 1:} For $1<R_0<1.5$,  the stationary PDF stays always unimodal. By increasing $\sigma^2/\bar{\lambda}$, the PDF peak drifts from the deterministic equilibrium $x_1$ towards zero. When the relative noise intensity $\sigma^2/\bar{\lambda}$ crosses the level marked by the blue curve in Fig.~\ref{fig:bifurcation_white}, the PDF mode is located at zero. Further increase of $\sigma^2/\bar{\lambda}$ results in more probability mass being accumulated at zero. In Figs. \ref{fig:pdfs_white}A, B, we show an example of this noise-induced transition, for $R_0=1.4$.
\item  \textit{Type 2:} For $1.5<R_0<2$, the PDF mode shifts towards zero as $\sigma^2/\bar{\lambda}$ increases, which is similar to the previous case.  However, in this case,  when $\sigma^2/\bar{\lambda}$ crosses the blue curve level, the PDF becomes bimodal, with the additional peak located at zero. By increasing $\sigma^2/\bar{\lambda}$ further, more probability mass accumulates at zero, and, after $\sigma^2/\bar{\lambda}$ crosses the level marked by the magenta curve in Fig.\ref{fig:bifurcation_white},  the PDF becomes unimodal at zero.  In Figs. \ref{fig:pdfs_white}C, D, we show an example of this noise-induced transition for $R_0=1.7$.
\item \textit{Type 3:}  For $R_0>2$,  PDF peak drift phenomenon has the opposite trend; by increasing $\sigma^2/\bar{\lambda}$, the PDF peak drifts towards higher values.  When $\sigma^2/\bar{\lambda}$ crosses the blue curve level, an additional PDF mode appears at zero, whose magnitude increases by further increase of $\sigma^2/\bar{\lambda}$.  In Figs. \ref{fig:pdfs_white}E, F,  we show an example of this noise-induced transition for $R_0=2.2$.
\end{itemize}
\begin{remark}[Disease severity predictions under white noise in contact rate]\label{rem:SIS_disease}
We observe that, by increasing noise levels, a PDF peak at zero appears eventually, making the eradication of disease more likely. 
However, for diseases with $R_0<2$ (corresponding to noise-induced transitions of types 1 and 2), the most likely final size of disease in the population, i.e., the non-zero mode at $x_m$, drifts towards zero, even for low white noise levels, for which no PDF peak at zero has appeared yet. This means that, for $R_0<2$, white noise in contact rate always results in less severe predictions for disease spread. Note that, many SIS-modeled diseases lie in the range of $1<R_0<2$, such as gonorrhea, $R_0=1.4$ \cite{Hethcote1984}, syphilis, $R_0=1.32-1.50$ \cite{Tsuzuki2018},  streptococcus pneumoniae (pneumococcus), $R_0=1.4$ \cite{Hoti2009}, tuberculosis, $R_0=1.78$ \cite{Zhao2017,Feng2005}. On the other hand, for highly contagious diseases with $R_0>2$ (e.g.  pertussis, $R_0=5.5$ \cite{Kretzschmar2010}) increase in noise levels results in more spread of the disease, since, in this case, the most likely endemic point $x_m$ drifts towards larger values.
\end{remark}

We note that bifurcation diagrams were also analyzed by M\'endez et al.~\cite{Mendez2012} for a stochastic SIS model slightly different than Eq.~\eqref{eq:SIS_stoch}. However, in \cite{Mendez2012}, only the Stratonovich solution was considered, the peak drift phenomenon was not studied, and a dimensionless parameter involving noise intensity $\sigma$ and curing rate $\gamma$ was chosen, instead of the more easily interpretable relative variance $\sigma^2/\bar\lambda$.

\subsection{SIS model under Ornstein--Uhlenbeck noise}\label{sec:SIS_OU}
In this section, we let the stochastic perturbation $\xi(t)$ to be the standard OU process $\xi^{\text{OU}}(t)$ with zero mean and autocorrelation~\eqref{eq:OU_cor}.
Recall that $\tau>0$ is the correlation time of the OU noise. For an SDE under OU excitation, we can approximate its stationary PDF by the equilibrium solution of a nonlinear Fokker--Planck equation which was only recently formulated~\cite{Mamis2019a,Mamis2021}.  For the case of stochastic SIS model~\eqref{eq:SIS_stoch} under OU noise, we are able to derive an approximate stationary PDF for the infected population fraction (see Appendix \ref{A:FP}). This stationary PDF is available in the explicit closed form,
\begin{figure}
	\centering
	\includegraphics[width=\textwidth]{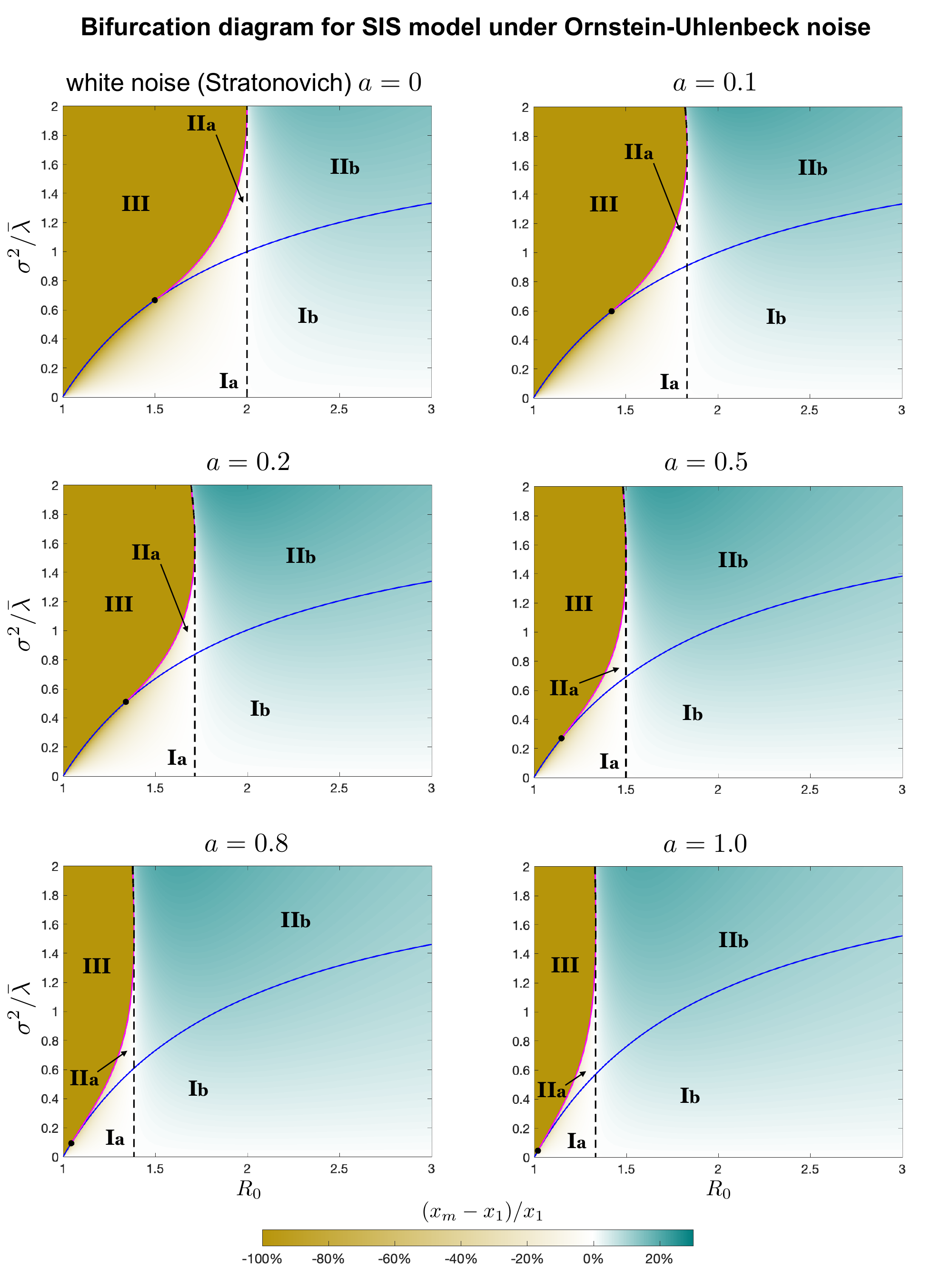}
	\caption{Bifurcation diagrams for the shape of stationary PDF of infected population fraction of stochastic SIS model \eqref{eq:SIS_stoch} under Ornstein--Uhlenbeck noise, for increasing relative correlation time $a=\tau/\eta$.  We define the depicted curves in Appendix \ref{A:extrema}. Regions and pseudocolor plot are the same as in Fig.~\ref{fig:bifurcation_white}.}\label{fig:bifurcation_OU}
\end{figure}

\begin{equation}\label{eq:p0_OU2}
p_0(x)=Cx^{Q_1}(1-x)(Gx^2-Dx+F)^{Q_2}\exp\left[Q_3\arctan\left(\frac{2Gx-D}{\sqrt{\vert \Delta\vert}}\right)\right],
\end{equation}
where $C$ is the normalization factor, and 
\begin{subequations}
\begin{equation}\label{eq:coeffs1}
Q_1=\frac{P}{F}-1, \ \ Q_2=-\frac{P}{2F}-1, \ \ Q_3=\frac{P(D-2BF)}{F\sqrt{\vert \Delta\vert}},
\end{equation}
\begin{equation}\label{eq:coeffs2}
G=A^2B^2+AB+1, \ \ D=A+AB+2A^2B+2, \ \ F=A^2+A+1,
\end{equation}
\end{subequations}
with $\vert\Delta\vert=4GF-D^2>0$, and 
\begin{equation}\label{eq:coeffs3}
P=\frac{2(1+a)}{B(\sigma^2/\bar{\lambda})}, \ \ B=\frac{R_0}{R_0-1}, \ \ A=\frac{a}{1+a}, \ \ a=\tau(\bar{\lambda}-\gamma).
\end{equation}
Despite its convoluted form, stationary PDF \eqref{eq:p0_OU2} depends on three dimensionless parameters only. Two of them, $R_0$ and $\sigma^2/\bar{\lambda}$, are the same as in the white noise case. The additional parameter $a=\tau/\eta$ is the \textit{relative correlation time} of the OU noise,  defined as the ratio of the correlation time $\tau$ of the noise and the Lyapunov characteristic time scale $\eta=(\bar{\lambda}-\gamma)^{-1}$ of the underlying deterministic model \eqref{eq:SIS}.  As discussed in Appendix \ref{A:FP}, for the white noise limit $\tau\rightarrow 0$, PDF \eqref{eq:p0_OU2} results in the Stratonovich stationary PDF \eqref{eq:p0_white} with $\varpi=1$.

Using PDF~\eqref{eq:p0_OU2}, we formulate the bifurcation diagrams shown in Fig.~\ref{fig:bifurcation_OU},  which depend on the dimensionless parameters $R_0$, $\sigma^2/\bar{\lambda}$ and $a$.  To the best of our knowledge, such bifurcation diagrams for the correlated noise case are considered here for the first time. 

\begin{figure}
	\centering
	\includegraphics[width=\textwidth]{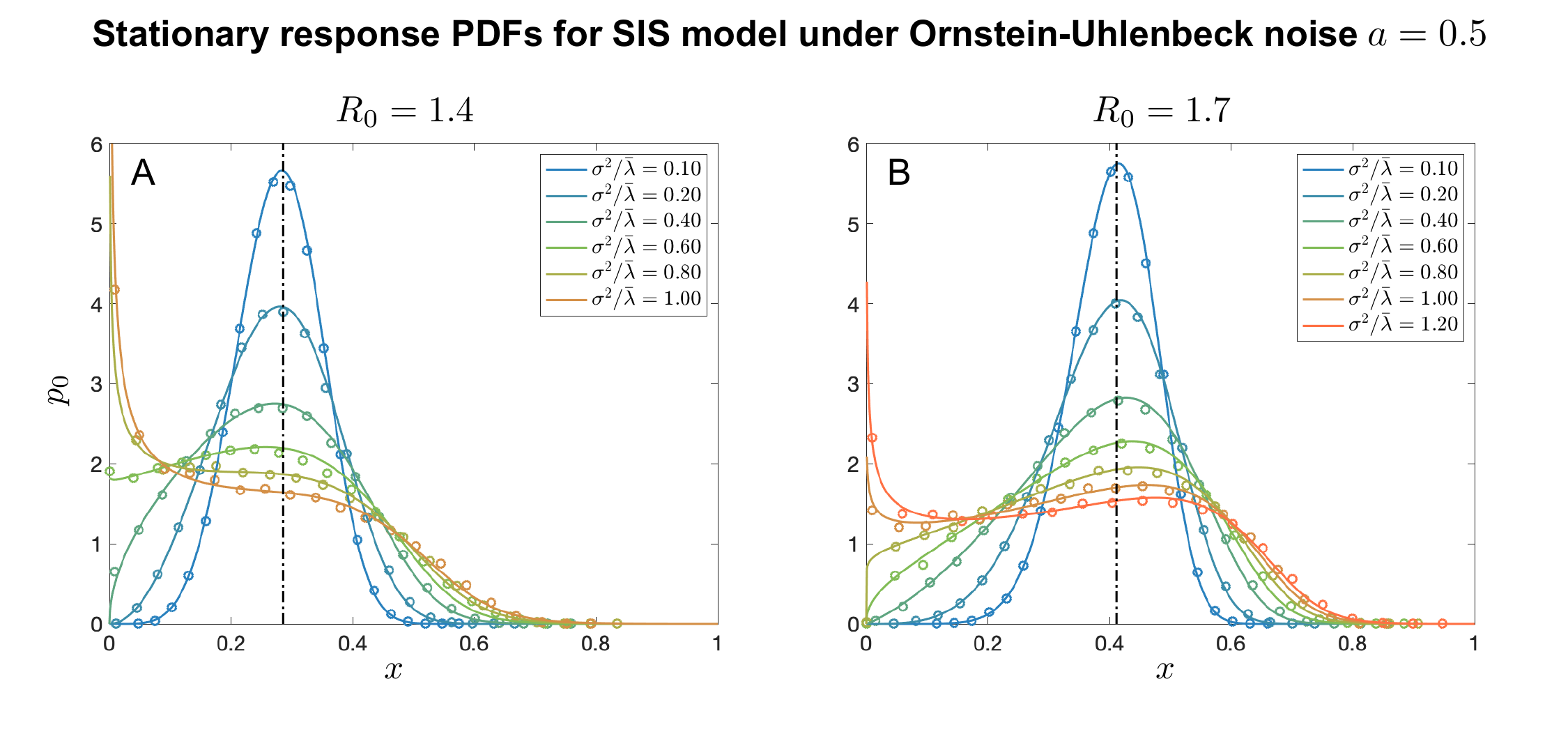}
	\caption{Stationary PDFs of infected population fraction of stochastic SIS model \eqref{eq:SIS_stoch} under Ornstein--Uhlenbeck noise with relative correlation time $a=0.5$,  for $R_0=1.4$ \textit{(panel A)}, and $R_0=1.7$ \textit{(panel B)}. In both figures, the deterministic equilibrium is denoted by a dash-dotted line,  the open circles correspond to PDFs obtained from 50,000 samples of direct Monte Carlo simulations, while solid lines mark the PDFs obtained from the analytic solution \eqref{eq:p0_OU2}. }\label{fig:pdfs_OU}
\end{figure}

As shown in Fig. \ref{fig:pdfs_OU}, although PDF \eqref{eq:p0_OU2} is approximate, it is in excellent agreement with the stationary PDFs obtained from direct Monte Carlo simulations of SDE \eqref{eq:SIS_stoch}.

Bifurcation diagrams in Fig.~\ref{fig:bifurcation_OU} corresponding to the correlated OU process are similar to those in Fig.~\ref{fig:bifurcation_white} for the uncorrelated white noise. In addition, the types of noise-induced transitions are similar to those in the white noise case. However, there are some important quantitative differences:
\begin{itemize}
\item Region III, where disease eradication is most likely, is smaller when using correlated OU process. Moreover, as the relative correlation time $a$ increases, this region shrinks further.
\item As the relative correlation time $a$ increases, the range of $R_0$ values corresponding to transitions of types 1 and 2 reduces. Furthermore, transitions of type 3 occur for $R_0$ that are significantly less than 2 (vertical dashed line).
\item PDF peak drift towards zero, that occurs in transitions of types 1 and 2, becomes less pronounced, as $a$ increases. 
\end{itemize}
To summarize, correlations in contact rate suppress the drift of the PDF mode towards zero and delay the emergence of a PDF mode at zero. This results in stationary PDFs whose probability mass is mainly located around the equilibrium of the deterministic SIS model. We can also observe the stabilizing effect of correlated noise by comparing Figs.~\ref{fig:pdfs_OU}A, B for OU noise with $a=0.5$, to the respective Figs.~\ref{fig:pdfs_white}B, D for white noise (Stratonovich interpretation).  We also observe the change in type of noise-induced transitions due to correlated noise:  for $R_0=1.4$ (resp., $R_0=1.7$), stationary PDF exhibits a type 1 (resp., type 2) noise-induced transition under white noise, while it exhibits a type 2 (resp., type 3) transition under OU noise with $a=0.5$.

\section{SEIR model}\label{sec:seir}
In this section, we consider the Susceptibles-Exposed-Infected-Removed (SEIR) model,
\begin{subequations}\label{eq:SEIR}
	\begin{equation}
		\frac{\id S(t)}{\id t}=-\frac{\lambda}{N}S(t)I(t),
	\end{equation}
	\begin{equation}
		\frac{\id E(t)}{\id t}=\frac{\lambda}{N}S(t)I(t)-\alpha E(t),
	\end{equation}
		\begin{equation}
		\frac{\id I(t)}{\id t}=\alpha E(t)-\gamma I(t),
	\end{equation}
			\begin{equation}\label{eq:Rem}
		\frac{\id R(t)}{\id t}=\gamma I(t).
	\end{equation}
\end{subequations}
Compared to the SIS model, the SEIR model has two additional compartments: the exposed $E(t)$ containing the individuals that have contracted the disease but are not infectious yet, and $R(t)$ containing the individuals that have been removed from the population, comprising the deceased and the immune due to vaccination or prior infection. The additional model parameter $\alpha$ is the average incubation rate, defined as the inverse of the average incubation (or latency) period during which the individual has contracted the disease but is not infectious yet.  SEIR models are suitable for describing the spread of airborne diseases such as flu and COVID-19, whose infection follows after a latency period, and also confers immunity after recovery, albeit temporarily \cite{Faranda2020a,Bertozzi2020,Piccolomini2020,Ghostine2021,Prabakaran2021, Antonelli2022}. 

In our study, we use SEIR model \eqref{eq:SEIR} to model the Omicron wave of COVID-19 pandemic in the US, i.e., the period between December 3, 2021 and April 22, 2022. We use the data for cumulative infections from the COVID-19 Dashboard by the Center for Systems Science and Engineering (CSSE) at Johns Hopkins University~\cite{JohnsHopkins, VargaLajos2020}. The total population $N$ is considered constant and equal to the US population of 329.5 million, and the initial values of the exposed $E(t_0)$, the infected $I(t_0)$, and the removed $R(t_0)$ at the beginning of the Omicron wave were chosen consistent with the Johns Hopkins data base to be 0.14\%, 0.18\% and 14.88\% of the total population, respectively. Then, SEIR model parameters $R_0=\lambda/\gamma$, $\alpha$ and $\gamma$ are determined by least square fitting so that the cumulative number of COVID-19 cases during Omicron wave, as predicted by the model, agrees with the Johns Hopkins data. To determine the cumulative number of COVID-19 cases from the SEIR model, we use the relation
\begin{equation}\label{eq:cumulative_inf}
\int_{t_0}^tI(s)\id s=\frac{1}{\gamma}(R(t)-R(t_0)),
\end{equation}
which is derived from Eq.~\eqref{eq:Rem}. By this process, we obtain the values $R_0=1.85$, $\alpha=1/3.5$ days$^{-1}$, $\gamma=1/1.2$ days$^{-1}$.

After fitting the deterministic SEIR model to data, we add noise fluctuations to the average contact rate. We consider both cases of white and OU noise,  for noise levels $0.5\leq \sigma^2/\bar\lambda\leq 3.0$.  We note that, in prior studies which use stochastic compartmental models for COVID-19, the parameter $\sigma/\bar\lambda$ is chosen to model the noise level~\cite{Mamis2022,Faranda2020a, Faranda2020, Rand1991}. However, here we use $\sigma^2/\bar\lambda$, since this is a dimensionless parameter. 

\begin{figure}
	\centering
	\includegraphics[width=0.95\textwidth]{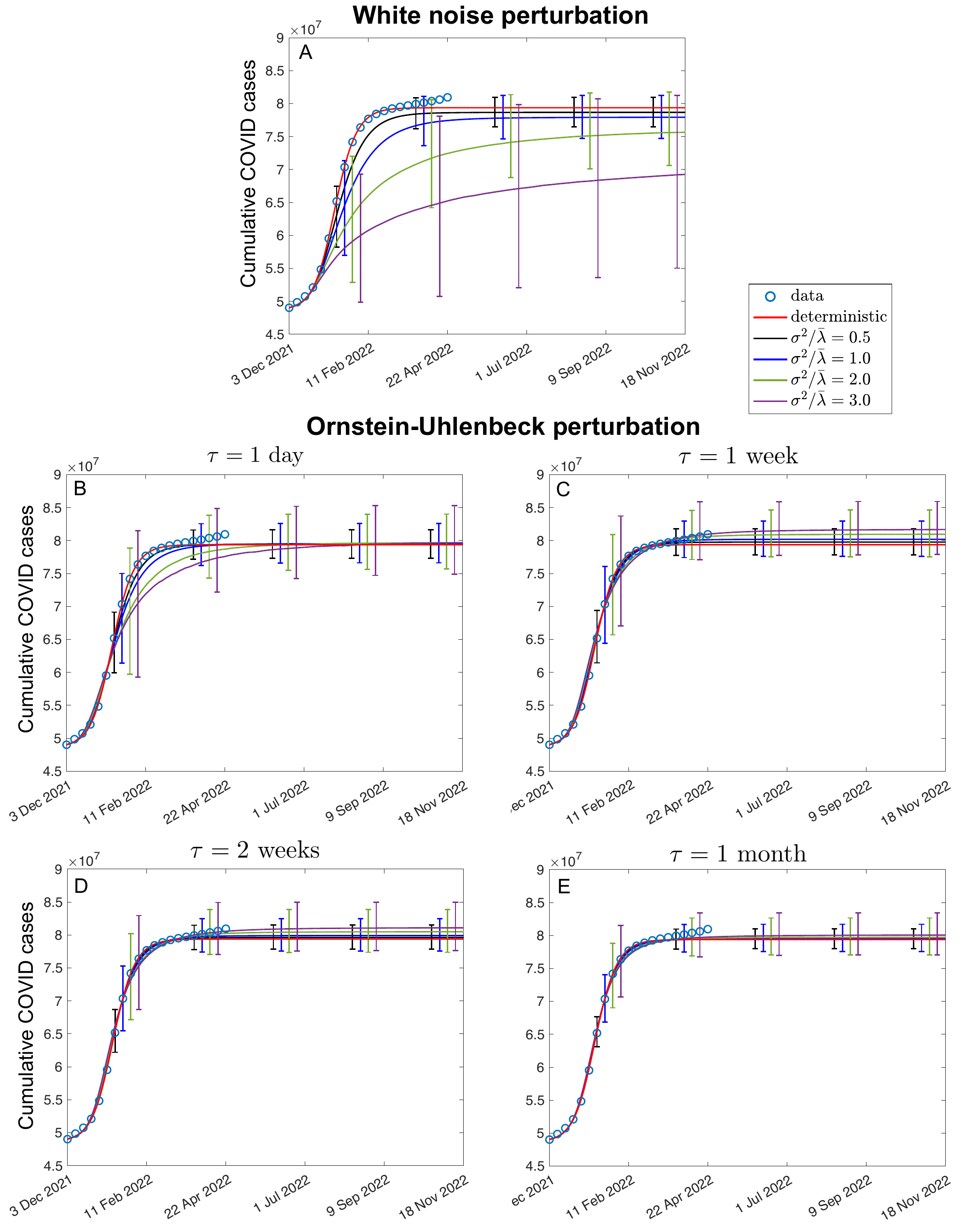}
	\caption{Predictions (mean trajectories with 50\% errorbars) of stochastic SEIR models for the cumulative number of COVID cases in the US during omicron wave, for increasing noise intensity.  \textbf{A:} white noise in contact rate. \textbf{B-E:} Ornstein-Uhlenbeck noise in contact rate with correlation time ranging from 1 day to 1 month.}\label{fig:seir_traj}
\end{figure}

\begin{figure}
	\centering
	\includegraphics[width=0.95\textwidth]{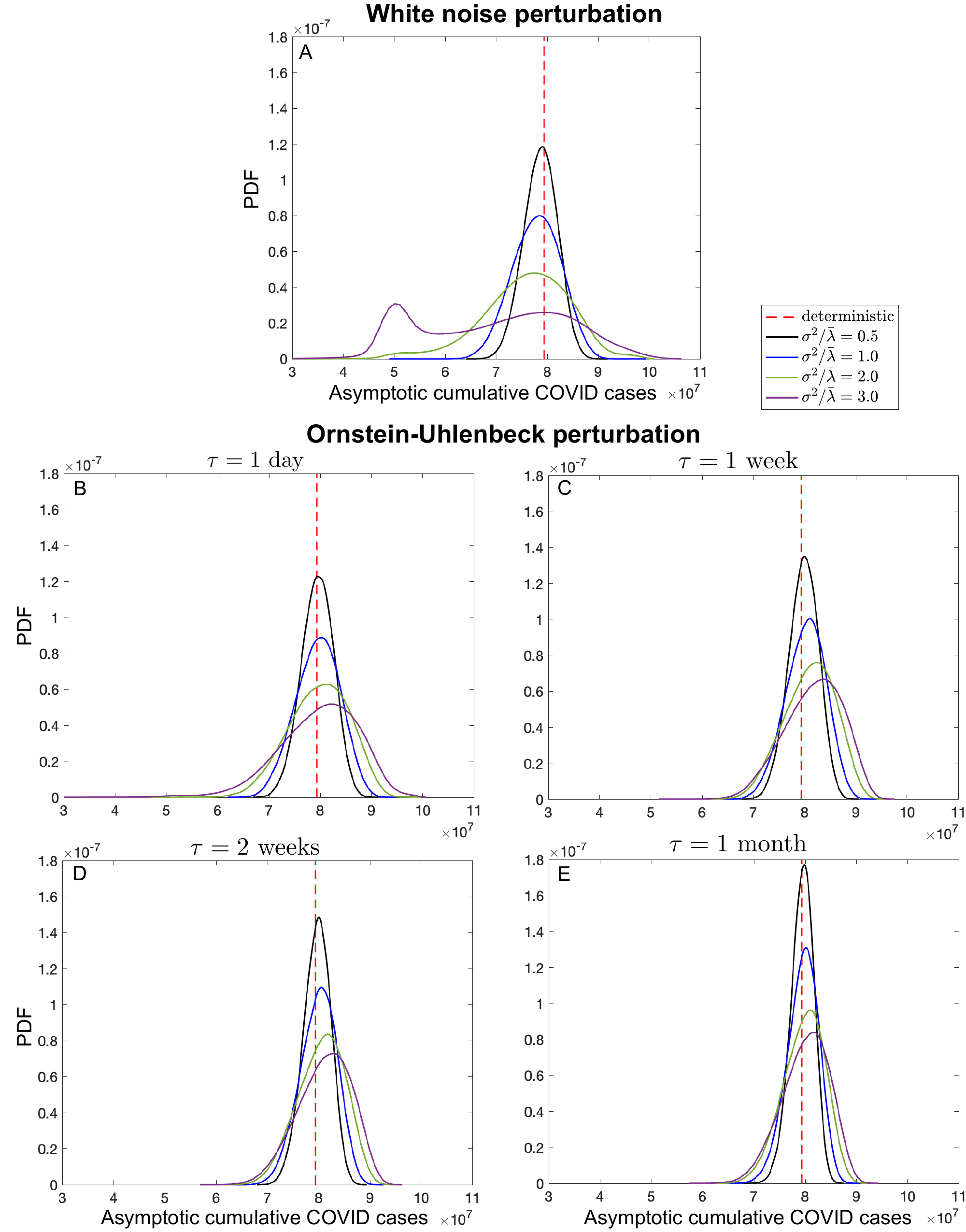}
	\caption{PDF of the asymptotic cumulative number of COVID cases in the US during omicron wave, for increasing noise intensity.  \textbf{A:} SEIR with white noise in contact rate. \textbf{B-E:} Ornstein--Uhlenbeck noise in contact rate with correlation time ranging from 1 day to 1 month. }\label{fig:seir_pdf}
\end{figure}

Contrary to the stochastic SIS model of Sec.~\ref{sec:det_SIS}, a stationary PDF for the stochastic SEIR model is not available in analytic form.  Thus, we perform Monte Carlo simulations of SEIR model~\eqref{eq:SEIR} with sample size 50,000 and stochastically perturbed contact rate $\lambda(t) = \bar\lambda+\sigma\xi(t)$.  In the white noise case, $\xi(t)=\xi^{\text{WN}}(t)$, the system of SDEs of SEIR model is numerically solved under the Stratonovich interpretation, using the predictor-corrector scheme of Cao et al.~\cite{Cao2015}. In the case of OU noise, $\xi(t)=\xi^{\text{OU}}(t)$, stochastic SEIR model is augmented by the linear SDE,
\begin{equation}\label{eq:OU_SDE2}
\frac{\id \xi^{\text{OU}}(t)}{\id t}=-\frac{1}{\tau}\xi^{\text{OU}}(t)+\frac{1}{\tau}\xi^{\text{WN}}(t),
\end{equation}
that generates the standard OU process $\xi^{\text{OU}}(t)$. The resulting coupled system is again solved using a predictor-corrector scheme~\cite{Cao2015}.
The time series of the mean cumulative COVID cases obtained from the Monte Carlo simulations are shown in Fig.~\ref{fig:seir_traj}. These simulations are also used to determine the stationary PDFs of COVID cases shown in Fig.~\ref{fig:seir_pdf}.

As we show in Fig.~\ref{fig:seir_traj}, the choice between white or OU noise for modeling uncertainties in the contact rate is consequential since they lead to very different forecasts for the spread of the pandemic. For increasing levels of white noise intensity, SEIR model significantly underestimates the severity of the pandemic on average. On the other hand, OU noise leads to forecasts whose mean trajectory of COVID cases stays always close to the actual data. The best fit is obtained for OU noise with correlation time of 1 week, which is in agreement with the weekly social patterns observed in human behavior \cite{Goh2008, Jiang2012, Kivela2012, Saramaki2015}.  Note however that, despite the abundance of data collected from the COVID-19 pandemic, the correlation time of the contact rate has not been quantified yet \cite{Tkachenko2021}.

The reason the SEIR model under white noise underestimates the pandemic spread is that it undergoes a noise-induced transition similar to type 2 transition of the stochastic SIS model, see Fig.~\ref{fig:seir_pdf}.  As the white noise intensity increases, the PDF peak drifts from the deterministic equilibrium towards lower values. Due to this peak drift, the mean trajectories of cumulative COVID cases for $\sigma^2/\bar\lambda=0.5$ and $\sigma^2/\bar\lambda=1$ lie below the actual data. For the noise level $\sigma^2/\bar\lambda=2$,  an additional peak emerges in the regime of low number of cases ($\approx5\times10^7$). Further increase of white noise intensity makes the additional peak more pronounced. This results in stochastic SEIR model to greatly underestimate the pandemic severity for $\sigma^2/\bar\lambda=3$. 

In contrast, when the contact rate is perturbed by the OU noise, the stationary PDF of COVID cases remains unimodal for a wider range of noise levels.  The presence of correlations in OU noise hinders the emergence of the additional peak at lower case values; only for the combination of small correlation time ($\tau=1$ day) and high intensity ($\sigma^2/\bar\lambda=3$) of the OU noise does an additional peak start forming around $5\times10^7$ cases (see Fig.~\ref{fig:seir_pdf}B). Also, the stationary PDF for OU noise exhibits the opposite trend in peak drift compared to the PDFs for white noise; increasing the noise intensity makes the peak to drift towards higher values.  Thus, presence of temporal correlation in the noise changes the type of the noise-induced transitions the SEIR model undergoes. This is similar to our results for correlated noise in the stochastic SIS model.

\begin{figure}
	\centering
	\includegraphics[width=0.95\textwidth]{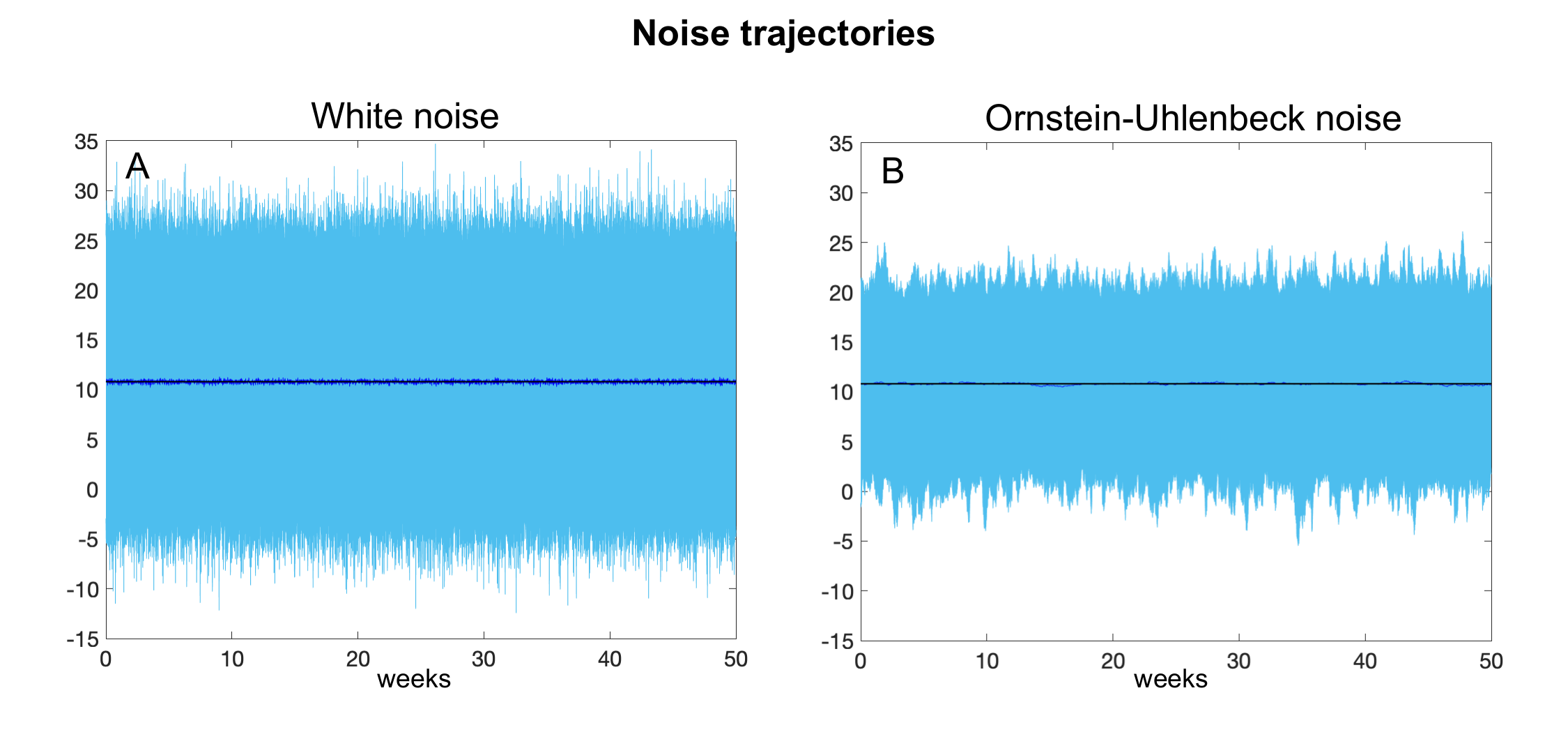}
	\caption{Envelope of 1,000 trajectories of the contact rate for COVID-19 perturbed around its mean value $\bar\lambda=10.8$ weeks$^{-1}$ (shown by black line) by noise of intensity $\sigma^2=2\bar\lambda$.  The mean of the 1,000 trajectories is shown by the blue curve. \textbf{A:} white noise perturbation \textbf{B:} OU perturbation, with correlation time $\tau=1$ week. }\label{fig:noise_plots}
\end{figure}

In Figs.~\ref{fig:seir_pdf}C-E, we also see that larger correlation times make the PDFs less diffusive, and the peak drift less pronounced.  This is the expected sharpening effect of correlated noise \cite{Hanggi1995, Mamis2021} as a result of the mean-reverting property of the OU process~\cite{Allen2016}, which becomes stronger as the correlation time increases.  The effect of mean-reverting property of OU noise is also shown in Fig.~\ref{fig:noise_plots}, where the OU noise is more concentrated around its mean value, compared to the white noise with the same intensity.  This also means that the OU noise becomes negative less frequently than the respective white noise.  Nonetheless, since OU noise is Gaussian and thus unbounded, it can always attain negative values, which is  unrealistic for the contact rate. Prior work on stochastic oncology remedies these unwanted negative values by considering bounded noise~\cite{Domingo2020,DOnofrio2010,DOnofrio2008}. 
Noise-induced transitions in compartmental models under bounded noise have not been studied extensively yet (see, e.g., \cite{Bobryk2005}), thus constituting an interesting direction for future work. 

\section{Conclusions}\label{sec:conclusions}
It was shown recently that time correlations are essential for modeling uncertainties in the contact rate of an infectious disease~\cite{Mamis2022}. Using standard results from the theory of stochastic processes, Mamis and Farazmand~\cite{Mamis2022} showed that the only feasible process for modeling such uncertainties is the Ornstein--Uhlenbeck (OU) process. However, to arrive at this conclusion, the authors assumed that the autocorrelation function of the contact rate has an exponentially decreasing form. In the present work, we proved the same result without making such onerous assumptions. Modeling the contacts of each individual as a Markov process, assuming a reasonable conditional probability for such contacts, and using the central limit theorem, we proved that the contact rate averaged over the population satisfies the Langevin equation corresponding to the OU process.

We studied the implications of this result on two typical examples of stochastic compartmental models in epidemiology; the SIS model which describes bacterial and sexually transmitted diseases, and the SEIR model which describes airborne diseases such as COVID-19.  Stochasticity enters into the compartmental models by considering stochastic fluctuations in the contact rate, to account for uncertainties in social behavior of individuals in the population. 

For the stochastic SIS model, we derived the exact stationary PDF of the infected population fraction for both cases of white and Ornstein--Uhlenbeck noise fluctuations in contact rate.  As a result, we were able to determine the noise-induced transitions that a stochastic SIS model undergoes, as well as the effect of temporal correlations in contact rate.  Our main result is that, for a range of $R_0$ corresponding to many SIS-modeled diseases (see Remark \ref{rem:SIS_disease}) white noise in contact rate makes the eradication of the disease more likely. This is an unrealistic behavior since greater uncertainty in measuring a model parameter should not lead to the eradication of the disease.  On the other hand,  the inclusion of correlations has a stabilizing effect on the stationary PDF of the infected population fraction, mitigating the unrealistic transitions towards zero infected population. 

The results for noise-induced transitions of stochastic SEIR models are similar to those for SIS models. By performing Monte Carlo simulations of a SEIR model fitted to data from the Omicron wave of COVID-19 pandemic in the US, we observed that white noise models of the contact rate lead to systematic underestimation of the pandemic severity. On the other hand,  when the contact rate is modeled as an OU process, the predicted number of COVID cases is always close to the actual data.  An important direction for future work is to develop analytic tools for stochastic SEIR models, similar to those that we have already developed for SIS models.

Our work demonstrates that the inclusion of correlated uncertainties in compartmental models is a central component for a realistic stochastic model of disease spread. If overlooked, this would lead to unrealistic, less severe forecasts.  However, despite the abundance of data collected, especially during the COVID-19 pandemic, the intensity and temporal correlations of noise in compartmental model parameters have not been determined with precision.  This calls for more empirical studies that would systematically quantify the nature of uncertainties, and especially their correlation time, in the parameters of compartmental epidemiological models.

\paragraph{Acknowledgments} K.M. would like to acknowledge the hospitality of the  Department of Mathematics at North Carolina State University where most of this work was carried out when he was a postdoctoral associate in the research group of M.F.

\paragraph{Data availability} The data used in this work is available from COVID-19 Dashboard
by the Center for Systems Science and Engineering (CSSE) at Johns Hopkins University,
\url{https://github.com/CSSEGISandData/COVID-19}.

\paragraph{Authors' contributions} M.F. conceptualized and supervised the research. K.M. conducted the
research and wrote the paper. M.F. revised the paper. 

\paragraph{Funding} The authors received no funding for this work.

\paragraph{Competing interests} The authors declare that they have no competing financial interests.

\begin{appendices}

\section{Calculation of stationary PDFs}\label{A:FP}
Consider the general form of the scalar SDE
\begin{equation}\label{eq:SDE_gen}
\frac{\id X(t)}{\id t}=h(X(t))+\sigma(X(t))\xi(t),
\end{equation}
where $X(t)$ and $\xi(t)$ are the stochastic processes of the response and excitation respectively,  $h(x)$ is the continuous drift function, and $\sigma(x)$ is the differentiable function of the noise intensity.  

In the case where excitation $\xi(t)$ is Gaussian white noise (see Eq.~\eqref{eq:WN_cor}), the evolution of the PDF $p(x,t)$ of the response $X(t)$ is governed by the classical Fokker--Planck equation (see, e.g., \cite[Ch. 5]{Gardiner2004}):
\begin{equation}\label{eq:FP}
\frac{\partial p(x,t)}{\partial t}+\frac{\partial}{\partial x}\left[\left(h(x)+\frac{\varpi}{2}\sigma'(x)\sigma(x)\right)p(x,t)\right]=\frac{1}{2}\frac{\partial^2}{\partial x^2}\left[\sigma^2(x)p(x,t)\right].
\end{equation}
In Eq.~\eqref{eq:FP}, the drift term $h(x)$ is augmented by $(1/2)\sigma'(x)\sigma(x)$ which is the Wong-Zakai correction (see \cite{Sun2006,Mamis2016e}) modeling the difference between It\=o ($\varpi=0$) and Stratonovich ($\varpi=1$) interpretations of SDEs under multiplicative white noise excitation.  The stationary solution $p_0(x)=\lim_{t\rightarrow\infty}p(x,t)$ of Fokker--Planck Eq.~\eqref{eq:FP} is given in the closed form \cite[Sec. 5.3.3]{Gardiner2004}:
\begin{equation}\label{eq:p0}
p_0(x)=\frac{C}{\sigma^{2-\varpi}(x)}\exp\left(2\int^x\frac{h(y)}{\sigma^2(y)}\id y\right),
\end{equation}
where $\int^x\id y$ denotes the antiderivative, and $C$ is the normalization factor.

In our recent papers~\cite{Mamis2019a,Mamis2021}, we derived an approximate nonlinear Fokker--Planck equation corresponding to SDE~\eqref{eq:SDE_gen} under correlated excitation:
\begin{align}\label{eq:nFP}
\frac{\partial p(x,t)}{\partial t}+\frac{\partial}{\partial x}&\left\{\left[h(x)+\sigma'(x)\sigma(x)A(x,t;p)\right]p(x,t)\right\}=\nonumber\\&=\frac{\partial^2}{\partial x^2}\left[\sigma^2(x)A(x,t;p)p(x,t)\right],
\end{align}
where
\begin{equation}\label{eq:A_M}
A(x,t;p)=\sum_{m=0}^2\frac{D_m(t;p)}{m!}\left\{\zeta(x)-\mathsf{E}[\zeta(X(t))]\right\}^m, 
\end{equation}
with 
\begin{equation}\label{eq:zeta}
\zeta(x)=\sigma(x)\left(\frac{h(x)}{\sigma(x)}\right)',
\end{equation}
and
\begin{equation}\label{eq:D_m}
D_m(t;p)=\int_{t_0}^tC_{\xi}(t,s)\exp\left(\int_s^t\mathsf{E}[\zeta(X(u))]\id u\right)(t-s)^m\id s.
\end{equation}
where $C_{\xi}(t,s)$ is the autocorrelation function of noise excitation $\xi(t)$.
Fokker--Planck equation \eqref{eq:nFP} is nonlinear, due to the dependence of coefficient $A(x,t;p)$ on the response moment $\mathsf{E}[\zeta(X(t))]$, which in turn depends on the unknown PDF $p(x,t)$.  As we have proven in \cite{Mamis2021}, for diminishing correlation time of $\xi(t)$, $\tau\rightarrow0$, coefficient $A$ in Eq.~\eqref{eq:nFP} becomes $1/2$. Thus,  in the white noise limit, nonlinear Fokker--Planck Eq.~\eqref{eq:nFP}  becomes the Stratonovich Fokker--Planck Eq.~\eqref{eq:FP} for $\varpi=1$.  Also, note that, by keeping only the zeroth-order term in the sum of Eq.~\eqref{eq:A_M}, we obtain the widely-used H\"anggi's approximate Fokker--Planck equation \cite{Hanggi1985,Hanggi1995}.

For $\xi(t)$ being the standard OU process (see Eq.~\eqref{eq:OU_cor}), the stationary solution of the nonlinear Fokker--Planck Eq.~\eqref{eq:nFP} reads
\begin{equation}\label{eq:p0_OU}
p_0(x,M)=\frac{C}{\sigma(x)A(x,M)}\exp\left(\int^x\frac{h(y)}{\sigma^2(y)A(y,M)}\id y\right),
\end{equation}
where $A(x,M)$ is the stationary value of coefficient $A(x,t;p)$, given by
\begin{equation}\label{eq:A_M_stat}
A(x,M)=\frac{1}{2}\sum_{m=0}^2\frac{[\tau(\zeta(x)-M)]^m}{(1-\tau M)^{m+1}},
\end{equation}
and $M$ is the stationary value of response moment $\mathsf{E}[\zeta(X(t))]$:
\begin{equation}\label{eq:R}
M=\int_{\mathbb{R}}\zeta(x)p_0(x,R)\id x.
\end{equation}
As derived in \cite{Mamis2021},  solution \eqref{eq:p0_OU} is valid under the condition $M<\tau^{-1}$.  

Due to the presence of the unknown response moment $M$, Eq.~\eqref{eq:p0_OU} is an implicit stationary solution of the nonlinear Fokker--Planck Eq.~\eqref{eq:nFP}.  In \cite{Mamis2021}, we proposed an iteration scheme for the calculation of $M$, by substituting the implicit form \eqref{eq:p0_OU} for $p_0(x,M)$ into the definition relation \eqref{eq:R} for $M$. The initial value of moment $M$ for the iteration scheme is calculated from the corresponding Stratonovich Fokker--Planck equation.  Implicit closed-form solution \eqref{eq:p0_OU}, supplemented by the iteration scheme for $M$, constitutes a semi-analytic form for the stationary response PDF for SDE~\eqref{eq:SDE_gen} under OU stochastic excitation.

However, for the special case of stochastic SIS model \eqref{eq:SIS_stoch}, we are able to calculate moment $M$ analytically.  Note that the calculation of moment $M$ in explicit closed form is, in general, not possible.  
\begin{lemma}[Calculation of stationary response moment $M$]\label{lem:M}
For stochastic SIS model \eqref{eq:SIS_stoch}, moment $M$ defined by Eq.~\eqref{eq:R}, is 
\begin{equation}\label{eq:M2}
M=-(\bar{\lambda}-\gamma).
\end{equation}
\end{lemma}
\begin{proof}
For stochastic SIS model,~\eqref{eq:SIS_stoch}, and by substituting Eq.~\eqref{eq:p0_OU} into Eq.~\eqref{eq:R}, the definition relation for $M$ is specified as
\begin{equation}\label{eq:M3}
M=C\int_{\mathbb{R}}\frac{\zeta(x)}{\sigma(x)A(x,M)}\exp\left(\int^x\frac{h(y)}{\sigma^2(y)A(y,M)}\id y\right)\id x,
\end{equation}
where $h(x)=\bar{\lambda}x(1-x)-\gamma x$, $\sigma(x)=\sigma x(1-x)$ and $\zeta(x)=-\gamma x/(1-x)$. 
By performing integration by parts, we obtain
\begin{align}\label{eq:M4}
M&=C\int_{\mathbb{R}}\frac{\zeta(x)\sigma(x)}{h(x)}\left[\exp\left(\int^x\frac{h(y)}{\sigma^2(y)A(y,M)}\id y\right)\right]'\id x=\nonumber\\
&=-C\int_{\mathbb{R}}\left(\frac{\zeta(x)\sigma(x)}{h(x)}\right)'\exp\left(\int^x\frac{h(y)}{\sigma^2(y)A(y,M)}\id y\right)\id x=\nonumber\\
&=C\sigma\gamma(\bar{\lambda}-\gamma)\int_{\mathbb{R}}\frac{1}{[\bar{\lambda}(1-x)-\gamma]^2}\exp\left(\int^x\frac{h(y)}{\sigma^2(y)A(y,M)}\id y\right)\id x.
\end{align}
On the other hand,  normalization factor $C$ of $p_0$ is defined as
\begin{equation}\label{eq:N}
C^{-1}=\int_{\mathbb{R}}\frac{1}{\sigma(x)A(x,M)}\exp\left(\int^x\frac{h(y)}{\sigma^2(y)A(y,M)}\id y\right)\id x,
\end{equation}
and after integration by parts:
\begin{align}\label{eq:N2}
C^{-1}&=\int_{\mathbb{R}}\frac{\sigma(x)}{h(x)}\left[\exp\left(\int^x\frac{h(y)}{\sigma^2(y)A(y,M)}\id y\right)\right]'\id x=\nonumber\\&=-\int_{\mathbb{R}}\left(\frac{\sigma(x)}{h(x)}\right)'\exp\left(\int^x\frac{h(y)}{\sigma^2(y)A(y,M)}\id y\right)\id x=\nonumber\\&
=-\sigma\gamma\int_{\mathbb{R}}\frac{1}{[\bar{\lambda}(1-x)-\gamma]^2}\exp\left(\int^x\frac{h(y)}{\sigma^2(y)A(y,M)}\id y\right)\id x.
\end{align}
By substituting Eq.~\eqref{eq:N2} into Eq.~\eqref{eq:M4}, we obtain Eq.~\eqref{eq:M2}.
\end{proof}

Using Eq.~\eqref{eq:M2}, we calculate coefficient $A$ to
\begin{equation}\label{eq:A_n}
A(x)=\frac{1}{2(1+a)}\sum_{m=0}^2\left(\frac{a}{1+a}\right)^m(1-x)^{-m}\left(1-\frac{R_0x}{R_0-1}\right)^m, 
\end{equation}
where $a=\tau(\bar{\lambda}-\gamma)>0$.  By having coefficient $A(x)$ in explicit form, we can perform the integration in Eq.~\eqref{eq:p0_OU} analytically,  obtaining thus the expression \eqref{eq:p0_OU2} for the stationary response PDF for SIS model \eqref{eq:SIS_stoch} under OU noise.

\section{Analysis of stationary PDFs for SIS models}\label{A:extrema}
\subsection{White noise model - It\=o solution}\label{A:Ito}
In the vicinity of zero,  response PDF \eqref{eq:p0_white} for $\varpi=0$ is $p_0(x)\sim x^{\frac{2\left(1-R_0^{-1}\right)}{\left(\sigma^2/\bar{\lambda}\right)}-2}$.  For
\begin{equation}\label{eq:green_curve}
\frac{2\left(1-R_0^{-1}\right)}{\left(\sigma^2/\bar{\lambda}\right)}-2<-1\Rightarrow (\sigma^2/\bar{\lambda})>2\left(1-R_0^{-1}\right),
\end{equation}
$p_0(x)$ is not integrable, and $p_0(0)=+\infty$. Thus, under condition \eqref{eq:green_curve}, $p_0(x)$ is a delta function at zero.  Eq.~\eqref{eq:green_curve} always holds true for $R_0<1$, resulting in disease eradication, as in the deterministic case.  This is the reason for choosing $R_0\in[1,+\infty)$ in our analysis. The green curve in Fig.~\ref{fig:bifurcation_white}A corresponds to $(\sigma^2/\bar{\lambda})=2(1-1/R_0)$.  For
\begin{equation}\label{eq:blue_curve1}
-1<\frac{2\left(1-R_0^{-1}\right)}{\left(\sigma^2/\bar{\lambda}\right)}-2<0\Rightarrow 1-R_0^{-1}<(\sigma^2/\bar{\lambda})<2\left(1-R_0^{-1}\right),
\end{equation}
$p_0(x)$ is integrable, and has a peak at zero.  The blue curve in Fig.~\ref{fig:bifurcation_white}A corresponds to $(\sigma^2/\bar{\lambda})=1-1/R_0$.  For
\begin{equation}
\frac{2\left(1-R_0^{-1}\right)}{\left(\sigma^2/\bar{\lambda}\right)}-2>0\Rightarrow (\sigma^2/\bar{\lambda})<1-R_0^{-1},
\end{equation}
$p_0(x)$ is integrable, and $p_0(0)=0$.  For the case $(\sigma^2/\bar{\lambda})<2\left(1-1/R_0\right)$, where $p_0(x)$ is integrable, its local extrema points for $x\in(0,1)$ are specified, by the first derivative test, as the roots of quadtratic equation
\begin{equation}\label{eq:quadratic}
2\left(\sigma^2/\bar{\lambda}\right)x^2+\left[1-3\left(\sigma^2/\bar{\lambda}\right)\right]x+\left(\sigma^2/\bar{\lambda}\right)+R_0^{-1}-1=0.
\end{equation}
The requirement of nonnegative discriminant results in the condition
\begin{equation}\label{eq:magenta_curve1}
R_0\geq\frac{8\left(\sigma^2/\bar{\lambda}\right)}{\left[1+\left(\sigma^2/\bar{\lambda}\right)\right]^2}.
\end{equation} 
Eq.~\eqref{eq:magenta_curve1}, for the case of equality, is the magenta curve in Fig. \ref{fig:bifurcation_white}A.  The two roots of Eq.~\eqref{eq:magenta_curve1} are
\begin{equation}\label{eq:roots}
x_{\pm}=\frac{3\left(\sigma^2/\bar{\lambda}\right)-1\pm\sqrt{\left[1+\left(\sigma^2/\bar{\lambda}\right)\right]^2-8\left(\sigma^2/\bar{\lambda}\right)R_0^{-1}}}{4\left(\sigma^2/\bar{\lambda}\right)}.
\end{equation}
By the additional requirement of $x_{\pm}\in(0,1)$,  we summarize the conditions for roots $x_{\pm}$ to be extrema points of $p_0(x)$.
\begin{equation*}
x_+\text{ is extremum point for}\left\{\begin{matrix}(\sigma^2/\bar{\lambda})<1/3\bigwedge(\sigma^2/\bar{\lambda})<1-1/R_0,\\(\sigma^2/\bar{\lambda})>1/3\bigwedge R_0\geq\frac{8\left(\sigma^2/\bar{\lambda}\right)}{\left[1+\left(\sigma^2/\bar{\lambda}\right)\right]^2}.\end{matrix}\right.
\end{equation*}
\begin{equation*}
x_-\text{ is extremum point for }(\sigma^2/\bar{\lambda})>1/3\textstyle{\bigwedge}(\sigma^2/\bar{\lambda})>1-1/R_0\textstyle{\bigwedge}R_0\geq\frac{8\left(\sigma^2/\bar{\lambda}\right)}{\left[1+\left(\sigma^2/\bar{\lambda}\right)\right]^2}.
\end{equation*}
Furthermore, we determine that $x_+$ is a maximum point, and $x_-$ is a minimum point. Thus, we identify $x_+$ as the non-zero mode coordinate $x_m$.  By using de l'H\^opital rule,  we calculate $\lim_{\sigma\rightarrow0}x_m=(\bar{\lambda}-\gamma)/\bar{\lambda}$, which is the expected result that, in the deterministic limit,  PDF mode $x_m$ coincides with the deterministic equilibrium.  

Last, in order to capture the peak drift phenomenon,  we calculate the first derivative of $x_m$ with respect to $(\sigma^2/\bar{\lambda})$. After algebraic manipulations, we obtain that
\begin{equation}\label{eq:peak_drift}
x_m'(\sigma^2/\bar{\lambda})\geq0\Rightarrow R_0\geq2.
\end{equation}
The dashed line in Fig.~\ref{fig:bifurcation_white}A is $R_0=2$.

\subsection{White noise model - Stratonovich solution}\label{A:Str}
We repeat the procedure we followed in Sec.~\ref{A:Ito}, for Stratonovich solution, Eq.~\eqref{eq:p0_white} for $\varpi=1$. The results we obtain are the following: Stratonovich solution is a delta function at zero only for $R_0<1$; for $R_0>1$, it is always integrable.  For
\begin{equation}\label{eq:blue_curve2}
(\sigma^2/\bar{\lambda})>2(1-1/R_0),
\end{equation}
$p_0(x)$ has a peak at zero.  The blue curve in Fig.~\ref{fig:bifurcation_white}B corresponds to $(\sigma^2/\bar{\lambda})=2(1-1/R_0)$.   For $x\in(0,1)$, the local extrema are roots of the equation
\begin{equation}\label{eq:quadratic2}
2\left(\sigma^2/\bar{\lambda}\right)x^2+\left[2-3\left(\sigma^2/\bar{\lambda}\right)\right]x+\left(\sigma^2/\bar{\lambda}\right)+2(R_0^{-1}-1)=0.
\end{equation}
Thus, the possible extrema points in $(0,1)$ are
\begin{equation}\label{eq:roots2}
x_{\pm}=\frac{3\left(\sigma^2/\bar{\lambda}\right)-2\pm\sqrt{\left[2+\left(\sigma^2/\bar{\lambda}\right)\right]^2-16\left(\sigma^2/\bar{\lambda}\right)R_0^{-1}}}{4\left(\sigma^2/\bar{\lambda}\right)},
\end{equation}
under the condition for nonnegative discriminant of Eq.~\eqref{eq:quadratic2}
 \begin{equation}\label{eq:magenta_curve2}
R_0\geq\frac{16\left(\sigma^2/\bar{\lambda}\right)}{\left[2+\left(\sigma^2/\bar{\lambda}\right)\right]^2}.
\end{equation} 
Eq.~\eqref{eq:magenta_curve2} for the case of equality is the magenta curve in Fig. \ref{fig:bifurcation_white}B.  We further determine that
\begin{equation*}
x_+\text{ is maximum point for}\left\{\begin{matrix}(\sigma^2/\bar{\lambda})<2/3\bigwedge(\sigma^2/\bar{\lambda})<2(1-1/R_0),\\(\sigma^2/\bar{\lambda})>2/3\bigwedge R_0\geq\frac{16\left(\sigma^2/\bar{\lambda}\right)}{\left[2+\left(\sigma^2/\bar{\lambda}\right)\right]^2}.\end{matrix}\right.
\end{equation*}
\begin{equation*}
x_-\text{ is minimum point for }(\sigma^2/\bar{\lambda})>2/3\textstyle{\bigwedge}(\sigma^2/\bar{\lambda})>2(1-1/R_0)\textstyle{\bigwedge}R_0\geq\frac{16\left(\sigma^2/\bar{\lambda}\right)}{\left[2+\left(\sigma^2/\bar{\lambda}\right)\right]^2}.
\end{equation*}
Also, we calculate that condition \eqref{eq:peak_drift} is true for Stratonovich solution too.

\subsection{Ornstein--Uhlenbeck noise model}\label{A:OU}
In the vicinity of zero,  response PDF \eqref{eq:p0_OU2} is $p_0(x)\sim x^{P/F-1}$. We calculate that, for $R_0<1$, solution \eqref{eq:p0_OU2} is a delta function at zero, similarly to the Stratonovich solution.  For
\begin{equation}\label{eq:blue_curve3}
\left(\sigma^2/\bar{\lambda}\right)>\frac{2(1+a)}{F}(1-R_0^{-1}),
\end{equation}
PDF \eqref{eq:p0_OU2} exhibits a peak at zero.  The blue curve in Fig.~\ref{fig:bifurcation_OU} corresponds to $\left(\sigma^2/\bar{\lambda}\right)=2(1+a)(1-R_0^{-1})/F$.  For $x\in(0,1)$, the local extrema are roots of the cubic equation $f(x)=0$, with
\begin{align}\label{eq:cubic}
&f(x)=2G\left(\sigma^2/\bar{\lambda}\right)x^3+\left[2(1+a)-(3G+D)\left(\sigma^2/\bar{\lambda}\right)\right]x^2+\nonumber\\
&\left[2D\left(\sigma^2/\bar{\lambda}\right)-2(1+a)(2-R_0^{-1})\right]x+2(1+a)(1-R_0^{-1})-F\left(\sigma^2/\bar{\lambda}\right).
\end{align}
The calculation of the exact roots of a cubic equation is cumbersome. However,   $f(1)=-\left(\sigma^2/\bar{\lambda}\right)A^2(B-1)^2<0$, $f(+\infty)=+\infty$, and thus, by intermediate value theorem,  cubic polynomial $f(x)$ has always a real root that is greater than 1, which is not admissible as extremum point of $p_0(x)$.  Thus, the regions III in Fig.~\ref{fig:bifurcation_OU} correspond to $\Delta_3<0$, where $\Delta_3$ is the discriminant of the cubic polynomial $f(x)$.  

Note also that $f(-\infty)=-\infty$, and $f(0)>0$ under the condition
\begin{equation}\label{eq:region_I}
\left(\sigma^2/\bar{\lambda}\right)<\frac{2(1+a)}{F}(1-R_0^{-1}).
\end{equation}
Using the intermediate value theorem again, we deduce that, under condition \eqref{eq:region_I}, polynomial $f(x)$ has three distinct real roots, with only one of them in $(0,1)$.  By combining this result to the behavior of PDF \eqref{eq:p0_OU2} at zero (see Eq.~\eqref{eq:blue_curve3}), we conclude that, under condition \eqref{eq:region_I}, PDF \eqref{eq:p0_OU2} is unimodal, with its mode at a non-zero $x_m$. 

Last, we observe that, for $a=0$,  condition \eqref{eq:blue_curve3} is identical to condition \eqref{eq:blue_curve2}, and the cubic polynomial $f(x)$ is factorized to
\begin{align}\label{eq:f_a0}
&f(x)=\nonumber\\&(x-1)\left\{2\left(\sigma^2/\bar{\lambda}\right)x^2+\left[2-3\left(\sigma^2/\bar{\lambda}\right)\right]x+\left(\sigma^2/\bar{\lambda}\right)+2(R_0^{-1}-1)\right\}.
\end{align}
We identify the second factor in the right-hand side of Eq.~\eqref{eq:f_a0} as the quadratic polynomial, whose roots detetermine the PDF extrema points in $(0,1)$ of the white noise case, under Stratonovich interpretation (see Eq.~\eqref{eq:quadratic2}). This finding shows the compatibility between results under OU noise with $a=0$ and the Stratonovich solution  for the white noise case. 

\end{appendices}

\bibliography{library}

\end{document}